\documentclass[journal]{IEEEtran}
\usepackage{enumitem}
\usepackage{latexsym}
\usepackage{cite}
\usepackage{ifpdf}
\usepackage{amsmath}
\usepackage{booktabs}
\usepackage{graphicx}
\usepackage{amsfonts}
\usepackage{multirow}
\usepackage{amssymb}
\usepackage{dsfont}
\usepackage{hyperref}
\usepackage{epsfig}
\usepackage{color}
\usepackage{colortbl}
\usepackage{setspace}
\usepackage{algorithm}
\usepackage{algpseudocode}
\usepackage[justification=centering]{caption}
\usepackage{tabulary}
\usepackage{multirow}
\usepackage{xurl}
\usepackage{threeparttable}
\definecolor{mygray}{gray}{.6}

\newcommand{\q}{|q|}

\mathchardef\mhyphen="2D
\newcommand{\Aa}{\mathcal{A}}

\newcommand{\Dd}{\mathcal{D}}

\newcommand{\Kk}{\mathcal{K}}

\newcommand{\Oo}{\mathcal{O}}

\newcommand{\Rr}{\mathcal{R}}

\newcommand{\Ss}{\mathcal{S}}
\newcommand{\Tt}{\mathcal{T}}

\newcommand{\Vv}{\mathcal{V}}

\newtheorem{theorem}{Theorem}
\newtheorem{defn}{Definition}[section]

\makeatletter
\def\ExtendSymbol#1#2#3#4#5{\ext@arrow 0099{\arrowfill@#1#2#3}{#4}{#5}}
\def\RightExtendSymbol#1#2#3#4#5{\ext@arrow 0359{\arrowfill@#1#2#3}{#4}{#5}}
\def\LeftExtendSymbol#1#2#3#4#5{\ext@arrow 6095{\arrowfill@#1#2#3}{#4}{#5}}
\makeatother

\newcommand\func[1]{\mathsf{#1}}

 

\begin{document}

\title{Prove You Owned Me: One Step beyond RFID Tag/Mutual Authentication}

\author{\IEEEauthorblockN{Shaoying Cai, %\IEEEauthorrefmark{1},
Yingjiu Li, %\IEEEauthorrefmark{2},
Changshe Ma, %\IEEEauthorrefmark{3},
Sherman S. M. Chow, %\IEEEauthorrefmark{4}, and
Robert H. Deng~\IEEEmembership{Fellow,~IEEE}} %\IEEEauthorrefmark{5},
%
%\IEEEauthorblockA{\IEEEauthorrefmark{1}College of Computer Science and Electronic Engineering, Hunan University, China}
%\IEEEauthorblockA{\IEEEauthorrefmark{2}Department of Information Engineering, The Chinese University of Hong Kong, Hong Kong}
%\IEEEauthorblockA{\IEEEauthorrefmark{3}School of Computer, South
%China Normal University, China}
% \IEEEauthorblockA{\IEEEauthorrefmark{4}Computer and Information Science Department, University of Oregon, USA}
% \IEEEauthorblockA{\IEEEauthorrefmark{5}School of Information Systems, Singapore Management University, Singapore}
\thanks{%Manuscript received December 1, 2012; revised August 26, 2015.
S. Cai is with the College of Computer Science and Electronic Engineering, Hunan University, China. 
Y. Li is with the Computer and Information Science Department, University of Oregon, USA.
C. Ma is with the School of Computer, South China Normal University, China.
S. Chow is with the Department of Information Engineering, The Chinese University of Hong Kong, Hong Kong.
R. H. Deng is with the School of Computing and Information Systems, Singapore Management University, Singapore.
}}

\markboth{}%Journal of \LaTeX\ Class Files,~Vol.~14, No.~8, August~2015%
{Shell \MakeLowercase{\textit{et al.}}: Bare Demo of IEEEtran.cls for IEEE Communications Society Journals}

\maketitle

%!TEX root = main.tex
\begin{abstract}

Radio Frequency Identification (RFID) is a key technology used in many applications.
In the past decades, plenty of secure and privacy-preserving RFID tag/mutual authentication protocols as well as formal frameworks for evaluating them have been proposed.
However, we notice that a property, namely proof of possession (PoP), has not been rigorously studied till now, despite it has significant value in many RFID applications. For example, in RFID-enabled supply chains, PoP helps prevent dishonest parties from publishing information about products/tags that they actually have never processed.

We propose the first formal framework for RFID tag/mutual authentication with PoP after correcting deficiencies of some existing RFID formal frameworks. Our framework is based on a new privacy notion--unp$^\#$-privacy, and a new security notion--PoP credential unforgeability.  
We provide a generic construction to transform an RFID tag/mutual authentication protocol  to one that supports PoP using a cryptographic hash function, a pseudorandom function (PRF) and a  
signature scheme. We prove that the constructed protocol is secure and privacy-preserving under our framework if %the underlying protocol is so.
all the building blocks possess desired security properties.
Finally, we show an RFID mutual authentication protocol with PoP.  
Arming tag/mutual authentication protocols with PoP is an important step to strengthen RFID-enabled systems as it bridges the security gap between physical layer and data layer, and reduces the misuses of RFID-related data.

\end{abstract}

%!TEX root = main.tex
\section{Introduction}
Radio Frequency Identification (RFID) technology has greatly facilitated collection and management of identification information in a wide range of applications, from supply chain and access management to stock tracing and payments. RFID systems consist of three main components: tags, readers, and backend servers. Tags are radio transponders attached to physical objects. Readers are radio transceivers that communicate with tags to identify or authenticate them based on information stored on backend servers. RFID technology enables automatic identification and information collection due to the wireless communication property. When combined with internet and networking technology, RFID-related information can be integrated, shared, and queried in real time.

The wireless communication property of RFID is a double-edge sword. Despite enhancing efficiencies and reducing costs on manpower, it also causes RFID-enabled systems to be vulnerable to a variety of attacks. An adversary may eavesdrop, replay, and manipulate RFID communications to obtain tag identifiers, track tag locations, impersonate RFID tags and RFID readers, and trigger denial of service without tag owners' awareness. Also, if an adversary compromises any RFID tags (e.g., via side-channel attack \cite{Agrawal02}), they may access all secret information stored on the tags.

Plenty of efforts have been devoted to securing communications between RFID readers and tags \cite{Li13}. Secure and privacy-preserving tag/mutual authentication is the most fundamental functionality to protect RFID systems against various attacks. RFID tags should be identified with assurance in the presence of attacks, and meanwhile without disclosure of any valuable information. Hundreds of RFID tag/mutual authentication protocols (e.g., \cite{Chien2017,Yeh10,Song08,Li07,Molnar04,Cai09,Juels05,Li2015,Wang2014,Ryu15,Shen2017,Zhang2014,Zhao14,Chiou18}) as well as dozens of formal frameworks (e.g., \cite{Avoine05,Burmester06,Vaudenay2007,Le2007,Juels07, Damgard08, Ng2008,Ha08,Paise08,Burmester09,Ma2009,Deng10,Canard10,Lai2010,Li2011,Deng2011,Hermans2011,Moriyama12,Su17,Yang2018}) for evaluating them have been proposed.  However, an indispensable property, proof of possession (PoP), which was briefly discussed in \cite{Grummt08}, has not been rigorously studied till now.

PoP is highly valuable to RFID applications in which information about real-world events related to RFID tags are stored for future use and/or shared over networks.  For example,
access management systems require that the logs of visiting events related to authenticated access cards be kept for  identifying suspicious visitors when anomalies  happen. In supply chain management, visibility event data related to authenticated tags may be shared among supply chain parties  through various platforms such as EPCglobal Network \cite{EPCDS,EPCglobal} and  blockchain-based product management platforms \cite{Cui2019,Xu2019}. These application scenarios require RFID systems to not only authenticate tags, but also  prove to other parties that they indeed have authenticated the tags.
%but also  provide PoP to prove to other parties that they indeed have authenticated the tags which they provide to or share with other parties.
Otherwise, the information about real-world events related to RFID tags may be manipulated even if the underlying tag/mutual authentication protocol works well. For example, malicious access system administrator may manipulate the logs of visiting events related to authenticated access cards, and dishonest supply chain parties may make up visibility data about certain tags/products without actually processing them.

We propose to extend tag/mutual authentication protocols to support PoP. Our major contributions are summarized below.

\begin{itemize}

 \item We study the existing formal frameworks for RFID tag/mutual authentication protocols. We refine Deng et al.'s RFID system model \cite{Deng10} to allow terminations during protocol executions. We correct Yang et al.'s claim \cite{Yang2018} about the relationship between two major RFID privacy notions,  unp$^*$-privacy  and ind-privacy, and discuss the deficiencies of their privacy notion, unp$^\tau$-privacy.

 \item We propose the first formal framework for RFID tag/mutual authentication with a new security notion, named PoP credential unforgeability, and a new privacy notion, named unp$^\#$-privacy. Unp$^\#$-privacy can be applied for analysing any RFID reader-tag communication protocols.

 \item We provide a generic construction to transform an RFID tag/mutual authentication protocol to additionally support PoP using a cryptographic hash function, a pseudo-random function (PRF), and a signature scheme.
       %We prove that the constructed protocol is secure and privacy-preserving under our framework if
%       all the building blocks possess desired security properties.
      We conduct formal security analysis of our construction, and then discuss its practicability.

 \item We refine a secure and unp$^*$-privacy-preserving RFID mutual authentication protocol, and then extend it to support PoP according to our generic construction. We prove that the refined protocol with PoP is secure and privacy-preserving under our framework. We also discuss its implementation.

  \end{itemize}

%!TEX root = main.tex
\section{Related work}

\subsection{RFID authentication protocols}

The existing RFID tag/mutual authentication protocols can be classified in two categories: symmetric key-based and PKC-based. With symmetric key-based protocols, a reader and a tag conduct unidirectional or bidirectional authentication based on some shared secrets. Current symmetric key-based protocols include cyclic redundancy code (CRC) checksum-based ones (e.g., \cite{Chien2017,Yeh10}), one-way hash function-based ones ( e.g., \cite{Song08,Li07,Molnar04,Cai09}), and symmetric encryption algorithms-based ones (e.g, \cite{Juels05}), to name a few. However, symmetric key-based tag/mutual authentication protocols inherently cannot support PoP.

 Elliptic curve cryptography (ECC) is the most lightweight PKC and has been shown to be applicable in resource-constrained RFID settings \cite{Tuyls06,Bu18}. Researchers have proposed many ECC-based RFID authentication protocols. % \cite{He15,Wang2014,Ryu15, Li2015, Shen2017,Kaur16, Zhang2014,Zhao14,Chiou18}.
 However, most of them are shown to be vulnerable \cite{He15}, and only a few  remain secure till now \cite{Zhang2014,Zhao14,Wang2014,Shen2017,Ryu15}.
We discover that some ECC-based protocols (e.g., \cite{Zhang2014,Shen2017,Zhao14}) are actually symmetric key-based in terms of tag authentication, thus cannot support PoP as well. Only a couple of protocols \cite{Wang2014,Ryu15} can be potentially extended to support PoP, but have not been further explored yet.

\subsection{RFID formal frameworks}

Formal RFID security and privacy frameworks are fundamental to the design and
analysis of robust RFID protocols.
In general, an RFID tag/mutual protocol should
satisfy (a) correctness, which means a valid tag/reader should always be accepted; (b) security, which means an invalid tag/reader should always be
rejected; and (c) privacy, which means tags should not be identified or traced by unauthorized entities. Till now, many formal RFID frameworks have been proposed (e.g., \cite{Avoine05,Burmester06,Vaudenay2007,Le2007,Juels07, Damgard08, Ng2008,Ha08,Paise08,Burmester09,Ma2009,Deng10,Canard10,Lai2010,Li2011,Deng2011,Hermans2011,Moriyama12,Su17,Yang2018}).
%\cite{Avoine05,Burmester06,Vaudenay2007,
%Le2007,Juels07,Damgard08,Ng2008,Ha08,Paise08,Burmester09,Ma2009,Deng10,Canard10,Lai2010,Li2011,Deng2011,Hermans2011,Moriyama12,Su17,Yang2018}.

Correctness and security definitions in existing RFID formal frameworks appear to be, to a large extent, equivalent. Among them, Deng et al.'s  \cite{Deng10,Deng2011}
 is
considered more elaborate than others \cite{Hermans2011}. It is full of subtleties in developing rigorous and precise privacy notions. Dozens of other  privacy notions have been proposed, and are systematically discussed in \cite{Su2012,Moriyama12,Coisel13,Moriyama14}. Below we briefly introduce typical RFID privacy notions.

%\begin{enumerate}
 \paragraph{Indistinguishability-based privacy notion} Intuitively, it requires that any adversary cannot distinguish
 two uncorrupted tags \cite{Avoine05, Juels07} or two groups of tags \cite{Hermans2011}.
 It is easy to apply for proving the privacy of protocols which are built with ind-secure primitives, such as an IND-CCA secure encryption scheme.
 %It is easy to show that an RFID protocol does not satisfy the indistinguishability-based notions with a counterexample.

 \paragraph{Unpredictability-based privacy notion}
 Intuitively, it requires that any adversary cannot distinguish protocol messages from random strings. The unpredictability-based privacy notions are easy to apply for proving the privacy of symmetric key-based protocols, which form the majority of existing RFID protocols. We will disucss more on unpredictability-based privacy notions \cite{Ha08,Ma2009,Lai2010,Li2011,Yang2018} in Section \ref{section:priNotion}.

 \paragraph{Vaudenay's privacy notion \cite{Vaudenay2007}} Intuitively, it requires that for any adversary, there exists a blinded adversary such that the advantage of the adversary to win the privacy game over the blinded one's is negligible, where the blinded adversary does not `use' the communication captured
during the protocol run in order to determine its output.
 Vaudenay defined the most comprehensive adversary types.
There are following works \cite{Ng2008,Paise08,Armknecht10} to consolidate adversary type, extend the definitions to address mutual authentication, and etc.

 \paragraph{Zero-knowledge-based privacy notion} Intuitively, it requires that whatever information an
adversary can obtain from interacting with a target tag, there exists a simulator who can provide indistinguishable similar information without interacting with the tag. Zk-privacy was proposed by Deng et al.~\cite{Deng10}.
Moriyama et al. \cite{Moriyama12} showed that zk-privacy is equivalent to ind-privacy \cite{Juels07} which was proposed by Juels and Weis.

 \paragraph{Universal composable (UC) model-based privacy notion} UC is a powerful notion proposed by Canetti \cite{Canetti01} to describe cryptographic protocols that behave like ideal functionality, and can be composed in arbitrary way. This is known as the strongest (computational) security model for
cryptographic protocols. Several UC-based frameworks have been proposed for achieving  RFID privacy \cite{Burmester06,Le2007, Burmester09,Su17}.

%\end{enumerate}

%!TEX root = main.tex
\section{Discussions on unpredictability-based privacy notions} \label{section:priNotion}

 Each category of privacy notion has its own advantages. The unpredictability-based privacy notions can be easily applied for analysing symmetric key-based protocols. These protocols rely on  relatively resource-friendly building blocks such as hash function and block cipher, and are suitable for low-cost RFID tags.

The first unpredictability-based privacy notion, called unp-privacy, was proposed by Ha et al. \cite{Ha08}, and further strengthened to unp'-privacy \cite{Ma2009}, then unp$^*$-privacy \cite{Li2011}, and finally unp$^\tau$-privacy \cite{Yang2018}.
In \cite{Yang2018}, Yang et al. %pointed out that und$^*$-privacy is still not strong enough. They
claimed that unp$^*$-privacy does not imply ind-privacy, which is in contrast to the previous belief that unp$^*$-privacy is stronger. We will show that their claim is not sound, and discuss  the deficiency  of unp$^\tau$-privacy.
 
\subsection{Unp$^*$-privacy}
We first briefly review the RFID system model and the adversary model of unp$^*$-privacy. An RFID system consists of a reader $R$ and a set of tags $\Tt$. An RFID tag/mutual authentication protocol contains three rounds.
 A reader first sends a challenge $c$ to a tag, then the tag responses with a message $r$, and finally the reader sends the last message $f$. $P_c$, $P_r$, and $P_{f}$ are $c$, $r$, and $f$'s message spaces respectively.

 An adversary is given access to the following oracles:
%
%The reader is considered secure, except that it can be activated to send a first round message. An adversary can conduct active attacks over the wireless reader-tag communication channel, and corrupt a tag to get its stored contents.

\begin{itemize}
 \item $O_1$: Upon queried, the reader initializes a session, and  returns $(sid,c)$.
 \item $O_2$: On inputs $(T_i, sid, c)$, it returns a message $r$.
 \item $O_3$: On inputs $(sid, c, r)$, it returns a message $f$.
 \item $O_4$: On an input $T_i$, it returns the tag $T_i$'s secret keys and internal state information.

\end{itemize}

Let $\Oo$ denote the set of the four oracles $\{O_1, O_2, O_3, O_4\}$ specified above. An adversary is a
$(t, n_1, n_2, n_3, n_4)$-adversary, if it makes oracle queries to $O_i$ without
exceeding $n_i$ times respectively, where $1 \leq i \leq 4$, and the running time is at most $t$.

We use the following notations. If $\Aa(\cdot,\cdot,\cdots)$ is a
 randomized algorithm, then $y \leftarrow \Aa(x_1, x_2, \ldots; \rho)$
 means that $y$ is assigned with the unique output of
 algorithm $\Aa$ on inputs $x_1$, $x_2$, $\ldots$ and coins $\rho$, while $y
 \leftarrow \Aa(x_1, x_2, \ldots)$ is a shorthand for first picking
 $\rho$ at random and then setting $y \leftarrow \Aa(x_1, x_2,\ldots)$.
$y \leftarrow \Aa^{O_1,\ldots, O_\upsilon}(x_1, x_2, \ldots)$ denotes that $y$
is assigned with the output of algorithm $\Aa$ which takes $x_1, x_2,
\ldots$ as inputs and has oracle accesses to $O_1, \ldots, O_\upsilon$. $\Pr[\emph{E}]$ denotes the probability that an
event $E$ occurs.

Now we introduce unp$^*$-privacy. Intuitively, achieving unp$^*$-privacy requires protocol transcripts to be unpredictable, and protocol execution results to be unobservable. The experiment ${\bf Exp}_{\Aa}^{\mathit{unp}^*}$ $[\kappa, l, n_1, n_2,n_3, n_4]$, denoted as ${\bf Exp}_{\Aa}^{\mathit{unp}^*}$ for short, is illustrated in Figure \ref{Figure:unpstar}.
Given the security parameter $\kappa$, an RFID system is set up with a reader $R$ and a set of $l$ tags, where $l$ is polynomial to $\kappa$. An adversary $\Aa$ can launch oracle queries without exceeding $n_1$, $n_2$, $n_3$, and $n_4$ overall calls to $O_1$, $O_2$, $O_3$, and $O_4$ respectively throughout the experiment. $\Aa$ consists of two algorithms, $\Aa_1$ and $\Aa_2$, which
run in two stages, the learning stage and the guess stage, respectively.
 In the learning stage, $\Aa_1$ queries the four oracles, and outputs an uncorrupted
challenge tag $T_c$ and state information $st$. Then the experiment chooses $b \in_R \{0, 1\}$. In the guess stage, if $b = 1$, the experiment forwards $\Aa_2$'s queries to the oracles and returns the results, so that $\Aa_2$ can really  interact with the reader and $T_c$;
else, the experiment returns random values from appropriate message spaces. Finally, $\Aa_2$ guesses $b$'s value and outputs $b'$. The experiment outputs 1 if $b'= b$, and outputs 0 otherwise.

\newsavebox{\unpstar}
\begin{lrbox}{\unpstar}
\begin{tabular}{|l|}
\hline\\[-0.8em]
 Experiment {\bf Exp$_{\Aa}^{\mathit{unp}^*}[\kappa, l, n_1, n_2, n_3, n_4]$}\\
 1. run \textsf{Setup($\kappa$)} to setup $(R,\Tt)$.\\
 ~~//learning stage \\
 2. $\{T_c, st\} \leftarrow \Aa_1^{\Oo}(R, \Tt)$. \\
 3. select $b \in_R \{0, 1\}$.  \\
 ~~//guess stage\\
 4. $b' \leftarrow \Aa_2^{O_1,O_2, O_3}(R, T_c, st)$; \\
 ~~~in this stage, when $\Aa_2$ queries $O_1$, $O_2$, and $O_3$, \\
 ~~~if $b = 1$, return the results from the oracles; \\
 ~~~else, return a random element from $P_c$, $P_r$, and $P_{f}$\\
 ~~~respectively.
 \\
 5. output 1 if $b'= b$; else, output 0.\\
\hline
\end{tabular}
\end{lrbox}

\smallskip
\begin{figure}[htbp]
\centering  \scalebox{0.9}
{\usebox{\unpstar}} \caption{Unp$^*$-Privacy Experiment
}\label{Figure:unpstar}
\end{figure}

\begin{defn}
The advantage of adversary $\Aa$ in the experiment $\mathbf{Exp}^{unp^*}_\Aa$
 is defined as:
\begin{eqnarray*}
% \nonumber to remove numbering (before each equation)
&&Adv^{unp^*}
_\Aa(\kappa, l, n_1, n_2, n_3, n_4) \\
 &=&|\Pr[\mathbf{Exp}^{unp^*}_\Aa(\kappa, l, n_1, n_2, n_3, n_4) = 1] - \frac{1}{2}|,
\end{eqnarray*}
where the probability is taken over the choice of the tag set $\Tt$ and the coin tosses of the adversary
$\Aa$.
\end{defn}

\begin{defn}
An adversary $\Aa$ $(\epsilon, t, n_1, n_2, n_3, n_4)$-breaks the unp$^*$-privacy of the RFID system
$(R, \Tt)$ if the advantage $Adv^{unp^*}(\kappa, l, n_1, n_2, n_3, n_4)$ of $\Aa$ in the experiment $\mathbf{Exp}^{unp^*}_\Aa$ is at least $\epsilon$,
 and the running time of $\Aa$ is at most $t$.
\end{defn}

\smallskip
\begin{defn}
[Unp$^*$-Privacy] An RFID system $(R,\Tt)$ is said to be $(\epsilon, t, n_1, n_2, n_3, n_4)$-unp$^*$-private,  if  for all sufficiently large $\kappa$ there exists no adversary who can $(\epsilon, t, n_1, n_2, n_3, n_4)$-break the unp$^*$-privacy of $(R,\Tt)$ for any $(\epsilon, t)$, where $t$ is polynomial in $\kappa$ and $\epsilon$ is non-negligible in $\kappa$.
\end{defn}

\subsection{Correction on the relation between ind-privacy and unp$^*$-privacy}

Li et al. proved that unp*-privacy implies ind-privacy \cite{Li2011}. However,
 Yang et al. claimed that unp$^*$-privacy does not imply ind-privacy \cite{Yang2018}. To support this claim, they provided a counterexample, formally proved that it satisfies unp$^*$-privacy, and then showed that it does not satisfy ind-privacy through a traceability attack. However, we discover that the counterexample does not satisfy unp$^*$-privacy in the first place.

 %Based on these findings, they While we find that this claim is not correct.
%
%
% pointed out that unp$^*$-privacy is still not strong enough.
%We will show that Yang et al.'s counterexample does not satisfy either unp$^*$-privacy or ind-privacy. And the traceability attack is conducted with an adversary more powerful than allowed in und$^*$-privacy and ind-privacy. Thus, the claim that unp$^*$-privacy does not imply ind-privacy cannot stand.

We review the counterexample first.
Let $F : \{0, 1\}^{l_k} \times \{0, 1\}^{l_d} \rightarrow \{0, 1\}^{l_r}$ be a PRF family, $ctr \in \{0, 1\}^{l_r}$
be a counter, and $pad \in \{0, 1\}^{l_{pad}}$ be a padding, where $l_c + l_{pad} = l_d$ and $l_c$ is the length of the challenge.
Each tag $T_i$ has a unique identity $\mathit{ID}_i$,
and is assigned with a secret key $k_i \in_R \{0, 1\}^{l_k}$. $T_i$ stores $k_i$, a counter $ctr_i$ with an initial value $1$, and a one-bit tag state $st_i$ with an initial value 0.
The protocol works as follows.
\begin{enumerate}
 \item The reader $R$ chooses $c \in_R\{0,1\}^{l_c}$ and sends it to $T_i$.

 \item Upon receiving $c$, the tag $T_i$ chooses $r_2 \in_R \{0, 1\}^{l_r}$ first. Then $T_i$ calculates $r_1 = F_{k_i}(c||pad) \oplus ctr_i$ if $st_i = 0$; else $r_1 = F_{k_i}(c||r_2) \oplus ctr_i$.\footnote{In the counterexample, some  inputs of $F$ are not with the length $l_d$. We do not correct these mistakes.} Finally, $T_i$ updates the counter as $ctr_i = ctr_i+1$ , sets $st_i = 1$, and sends $(r_1, r_2)$ to $R$.
 \item Upon receiving $(r_1, r_2)$ from $T_i$, the reader $R$ searches for the matching tag in its database. If $R$ discovers a tuple $(k, ctr, \mathit{ID})$ such that $ctr = F_{k}(c||pad) \oplus r_1$, then  accepts $T_i$ as the tag with $\mathit{ID}$. Then $R$ updates $ctr = ctr+1$, computes $f= F_{k}(c||ctr||r_2)$; else if there exists $(k, ctr, \mathit{ID})$ such that $ctr = F_{k}(c||r_2) \oplus r_1$, then $R$  accepts $T_i$ as the tag with $\mathit{ID}$, updates $ctr=ctr+1$, computes $f= F_{k}(c||ctr||r_2)$ ; or else, $R$ rejects $T_i$  and chooses $f \in_R\{0,1\}^{l_r}$. At last,   $R$ sends $f$ to  $T_i$.
 \item Upon receiving $f$, if $f= F_{k_i}(c||ctr_i||r_2)$, $T_i$ sets $st_i =0$ and accepts the
reader $R$; otherwise, rejects $R$.
\end{enumerate}

Now  we show that the counterexample does not satisfy unp$^*$-privacy. An adversary $\Aa$ can break unp$^*$-privacy as follows.
In the learning stage, $\Aa_1$ triggers a valid tag $T_i$ which is in state 0 to run a session with the reader without modifying the messages. To do so, $\Aa_1$ first calls $O_1$ and gets $(sid,c)$, then calls $O_2$ with inputs $(T_i,sid,c)$  and gets $r_1$, where $r_1 = F_{k_i}(c||pad) \oplus ctr_i$;
and finally calls $O_3$ with inputs $(sid, c, r)$ and gets $f$. After running the above session, $T_i$'s state $st_i$ is still 0. %Note that, $\Aa_1$ can turn any tag's state from 1 to 0 by triggering the reader to complete a session with the tag as shown above.
Now $\Aa_1$ submits $T_i$ as the target tag. After the experiment tosses the coin $b$, in the guess stage, $\Aa_2$ calls $O_2$ with inputs $(T_i,sid',c)$  and gets $r'$, where $sid'$ could be any value in the session ID's space. As known, if $b = 1$, the experiment sets $r' = F_{k_i}(c||pad) \oplus (ctr_i+1)$; else, sets $r' \in_R P_r$. With the knowledge of $F_{k_i}(c||pad) \oplus ctr_i$, $\Aa_2$ can differentiate $F_{k_i}(c||pad) \oplus (ctr_i+1)$ from  a random string. Thus $\Aa_2$ can successfully guess the value of $b$.

Yang et al. showed a traceability attack against the counterexample.
If the tag $T_i$'s state $st_i$ is $0$, the reader can authenticate $T_i$ with $r_1$  without checking $r_2$'s integrity.
Thus, an adversary can infer  the value of $st_i$ by modifying $r_2$ and observing the reader's  protocol execution result. If the reader $R$ accepts $T_i$, the adversary knows that the value of $st_i$ is 0; or else, the value of $st_i$ is 1.
This kind of traceability attack cannot be captured by unp$^*$-privacy due to two reasons. First, an adversary cannot access protocol execution results in the unp$^*$-privacy experiment. Second, the soundness experiment \cite{Li2011} for tag authentication does not require the transcripts of a session to be matching (as defined in \cite{Deng10}) on the reader side and the tag side.

 %Thus, this kind of traceability attack cannot be captured by unp$^*$-privacy, despite it is indeed practical. Thus it is necessary to have a stronger privacy notion.
 
\subsection{Unp$^\tau$-privacy}

To capture the above traceability attack, Yang et al. proposed a new privacy notion, called unp$^\tau$-privacy. The experiment ${\bf Exp}_{\Aa}^{\mathit{unp}^\tau}$ is illustrated in Figure \ref{Figure:unptau}. Intuitively, unp$^\tau$-privacy requires that protocol transcripts are pseudorandom, and any modification on the second/third message would result in ``rejection'' by the reader/tag.
%, which is stronger than unp$^*$-privacy
%For simplify, we do not review full definition of unp$^\tau$-privacy, but only briefly introduce its experiment ${\bf Exp}_{\Aa}^{\mathit{unp}^\tau}$ which is roughly similar to the unp$^*$-privacy experiment
%which is illustrated in Figure \ref{Figure:unptau}.
%The experiment procedures are roughly similar to the unp$^*$-privacy experiment.
% Compared with unp$^*$-privacy, in unp$^\tau$-privacy,
 Compared to unp*-privacy, unp$^\tau$-privacy is different in several ways.
First,  an adversary can access one additional oracle $O_5$.
On input $(sid, f, T_i)$, $O_5$ returns $o_{T_i}$ which is $T_i$'s execution result  of the session $sid$. Second,  without being  specified explicitly, an adversary can get the reader's protocol execution results through querying  $O_3$. Third,  in the guess stage, if $b=0$, the experiment provides not only random chosen ``protocol transcripts'', but also ``protocol execution results''. The experiment outputs an ``execution result'' as ``1'' when $O_3$/$O_5$ is queried with the output of its preceding oracle, $O_2$/$O_3$, without modification; or ``0'' otherwise. % For example, if $\Aa_2$ queries $O_3$ with inputs $(sid, r, c)$, and $r$ is the output of a previous query to $O_2$ with inputs $(T_c, sid, c)$, the experiment returns a value $f \in_R P_f$ and the reader's execution result as ``1''; or else, the experiment returns a value $f \in_R P_f$ and the reader's execution result as ``0''.

%
%We would like to discuss more on how the experiment ${\bf Exp}_{\Aa}^{\mathit{unp}^\tau}$ generates protocol execution results when $b = 0$. Upon querying $O_3$/$O_5$, the experiment only outputs ``accept'' when it is queried with the output of its preceding oracle, say $O_2$/$O_3$, without modification. For example, if $\Aa_2$ queries $O_3$ with inputs $(sid, r, c)$, and $r$ is the output of a previous query to $O_2$ with inputs $(T_c, sid, c)$, the experiment returns a value $f \in_R P_f$ and the reader's execution result as ``1''; or else, the experiment returns a value $f \in_R P_f$ and the reader's execution result as ``0''.
%That is, any modification on the second/third message should be detected and results in ``rejection'' by the reader/tag.

\newsavebox{\unptau}
\begin{lrbox}{\unptau}
\begin{tabular}{|l|}
\hline\\[-0.8em]
 Experiment {\bf Exp$_{\Aa}^{\mathit{unp}^\tau}[\kappa, l, n_1, n_2, n_3, n_4, n_5]$}\\
 1. run \textsf{Setup($\kappa$)} to setup $(R,\Tt)$.\\
 ~~//learning stage \\
 2. $\{T_c, st\} \leftarrow \Aa_1^{O_1,O_2,O_3,O_4,O_5}(R, \Tt)$. \\
 3. select $b \in_R \{0, 1\}$. \\
 ~~//guess stage\\
 4. $b' \leftarrow \Aa_2\{O_1,O_2,O_3, O_5\}(R, T_c, st)$; \\
 ~(1) if $b = 1$, when $\Aa_2$ queries $O_1$,$O_2$,$O_3$, and $O_5$,  \\
 ~~~~~return the results of the oracles. \\

 ~(2) if $b = 0$,  \\
 ~~~a. when $\Aa_2$ queries $O_1$ and $O_2$, return random \\
 ~~~~~~values from $P_c$ and $P_r$ respectively.  \\
 ~~~b. when $\Aa_2$ queries $O_3$ with inputs $(sid, c, r)$, \\
 ~~~~~ if $r$ is the output of $O_2(T_c,sid,c)$, \\
 ~~~~~~~ return $f \in_R P_f$  and $o_R  = 1$;\\
 ~~~~~ else, return $f \in_R P_f$ and $o_R  = 0$. \\
 ~~~c. when $\Aa_2$ queries $O_5$ with inputs $(sid,f)$, \\
 ~~~~~ if $f$ is the output of $O_3(sid,c,r)$  and $o_R  = 1$, \\
 ~~~~~~~ return $o_T  = 1$;\\
 ~~~~~ else return $o_T  = 0$. \\
 5. output 1 if $b' = b$,  and 0 otherwise.\\
\hline
\end{tabular}
\end{lrbox}
\begin{figure}[htbp]

\centering  \scalebox{0.9}
{\usebox{\unptau}} \caption{Unp$^\tau$-Privacy Experiment
}\label{Figure:unptau}
\end{figure}

\smallskip
\begin{defn}
The advantage of adversary $\Aa$ in the experiment $\mathbf{Exp}^{unp^\tau}_\Aa$
 is defined as:
\begin{eqnarray*}
% \nonumber to remove numbering (before each equation)
&&Adv^{unp^\tau}
_\Aa(\kappa, l, n_1, n_2, n_3, n_4) \\
 &=&|\Pr[\mathbf{Exp}^{unp^\tau}_\Aa(\kappa, l, n_1, n_2, n_3, n_4) = 1] - \frac{1}{2}|,
\end{eqnarray*}
where the probability is taken over the choice of the tag set $\Tt$ and the coin tosses of the adversary
$\Aa$.
\end{defn}

\smallskip

\begin{defn}
An adversary $\Aa$ $(\epsilon, t, n_1, n_2, n_3, n_4)$-breaks the unp$^\tau$-privacy of the RFID system
$(R, \Tt)$ if the advantage $Adv^{unp^\tau}(\kappa, l, n_1, n_2, n_3, n_4)$ of $\Aa$ in the experiment $\mathbf{Exp}^{unp^\tau}_\Aa$ is at least $\epsilon$,
 and the running time of $\Aa$ is at most $t$.
\end{defn}

\smallskip
\begin{defn}
[Unp$^\tau$-Privacy] An RFID system $(R,\Tt)$ is said to be $(\epsilon, t, n_1, n_2, n_3, n_4)$-unp$^\tau$-private,  if  for all sufficiently large $\kappa$ there exists no adversary who can $(\epsilon, t, n_1, n_2, n_3, n_4)$-break the unp$^\tau$-privacy of $(R,\Tt)$ for any $(\epsilon, t)$, where $t$ is polynomial in $\kappa$ and $\epsilon$ is non-negligible in $\kappa$.
\end{defn}

\smallskip

\emph{Discussions.}
 We notice several issues with the definition of  unp$^\tau$-privacy.  First, $O_3$ is too powerful. An adversary can query $O_3$ with arbitrarily selected inputs $(sid, c, r)$, and get a response $f$.
 This contradicts the common assumption that the reader's internal routine cannot be interfered by an adversary.
 Second, to satisfy unp$^\tau$-privacy, the reader is required to send a dummy third-round message $f$ even if it has rejected the tag after receiving an invalid second round message, which is obviously not necessary.
 Third,  no session management is defined in Exp$_{\Aa}^{\mathit{unp}^\tau}$. It is  not clear what will happen when an adversary queries an oracle with the same inputs multiple times, or with different inputs that share the same $sid$ multiple times.

%!TEX root = main.tex

\section{Formal framework for RFID system with PoP}

We propose the first formal framework for RFID tag/mutual authentication with PoP under a refined RFID system model. % Our framework contains a new privacy notion, named unp$^\#$-privacy, and a new security notion, named PoP credential unforgeability.
Our framework also addresses the above  issues of unp$^\tau$-privacy.

\subsection{System model} \label{section:model}

 Among the existing RFID system models for tag/mutual authentication, Deng et al.'s \cite{Deng10} is
 considered to be more generic and elaborate than others. However, their model requires a reader/tag to continue sending a reply instead of terminating the execution upon receiving an invalid message.
 We refine Deng et al.'s model to allow protocol terminations during executions.

\smallskip
\begin{defn} \label{def:rs}
 An RFID system $\mathit{RS}$ is defined to be a tuple $(R, \Tt, \func{Setup}, \pi)$, where
\end{defn}

\begin{itemize}
 \item $\func{Setup}(\kappa,l)$ initializes the whole system $\mathit{RS}$ with a single legitimate reader $R$ and $l$ tags $\Tt$ with public system parameter $\sigma$, where $l$ is polynomial in the security parameter $\kappa$.
 Each tag $T_i \in \Tt$, for $1 \leq i \leq l$, is assigned with an unique identity $\mathit{ID}_i$,   public parameters $\xi_{T_i}$, an initial secret key $k_{T_i}^1$, and an initial internal state $s_{T_i}^1 = \emptyset$. The reader $R$ is assigned with a secret key $k_{R}$, an initial internal state $s_R^1 = \emptyset$, and a database $\mathit{DB}^1$.\footnote{Although the reader's internal state and database will be updated continually, we assume that all the historical contents of them are stored and can be retrieved later.}
 For each tag $T_i$, the database stores a record $rcd_{T_i}$ which contains information for authenticating $T_i$ (e.g., $T_i$' secret key).
 $\func{Setup}(\kappa,l)$ also outputs the whole system's public parameters $para = \{\sigma, \xi_{T_1}, \ldots, \xi_{T_l}\}$.

\vspace{1mm}
  \item $\func{Protocol}$ $\pi(R, T_i)$, denoted as $\pi$ for short, is a  $(2 \gamma +1)$-round interactive protocol between $R$ and $T_i$.\footnote{For a protocol that contains $2\gamma$ round messages, we   simply set  the $(2\gamma+1)$-th round message as an empty string.} Each session of $\pi$ is initialized by the reader with a fresh session ID $sid$ randomly selected in its space  $P_{sid}$. $c_\mu$/$\alpha_\mu$ denotes the $\mu$-th  reader-to-tag/tag-to-reader message in a session, and  $P_{c_\mu}$/$P_{\alpha_\mu}$ denotes $c_\mu$/$\alpha_\mu$'s message space, for $1 \leq \mu \leq (\gamma+1)/\gamma$. $R$/$T_i$ terminates a session by outputting $o_R$/$o_{T_i}$ which is a one-bit value for showing its protocol execution result of the session. $o_R$/$o_{T_i}$ is set as ``1'' for acceptance, ``0'' for denial, and ``$\mathsf{null}$'' if the execution result is not available. $o_R$/$o_{T_i}$ can be sent through a channel different from the wireless channel used by protocol messages (e.g., sound channel). The transcripts of a  session are defined as $trs^{sid} = (sid, c_1, \alpha_1, \ldots , c_\gamma,$ $\alpha_\gamma, c_{\gamma+1}, o_R, o_{T_i})$.

 % The protocol $\pi$ is shown in Fig. \ref{fig:gp1}. We omit to show $sid$ in each message, and put $o_R$/$o_{T_i}$ in parentheses to demonstrate that it is sent at the same time with the corresponding message but may with a different channel (e.g., sound channel).

% \begin{figure}[!htbp]
% \centering
% % Requires \usepackage{graphicx}
% \includegraphics[scale=0.55]{Figures/gp1.png}\\
% \caption{An RFID Protocol}\label{fig:gp1}
% \end{figure}

\smallskip
We assume that both of the reader and the tags  process protocol sessions sequentially, and a tag can run the protocol without exceeding $s$ times in its lifetime.

 \smallskip
\textbf{(1) On reader side:}  Suppose the reader $R$ is with current state ${s^j_R}$, current database $\mathit{DB}^j$, for $1 \leq j \leq ls$; there are three cases.

\smallskip
 \textbf{\emph{Case 1}}: $R$ can start its $j$-th session if ${s^j_R} = \emptyset$. It takes $para$, $k_R$, ${s^j_R}$, and $\mathit{DB}^j$ as inputs, and generates  ${s^j_R}'$ which contains partial transcripts $trs^{sid}$ $= (sid, c_1)$ and other temporary information, e.g. random coins for this session, where $sid$ is a fresh session ID, and $c_1$ is the first message in this session.  Then $R$ updates ${s^j_R}$ to ${s^j_R}'$, and  sends $(sid, c_1) $ to $T_i$.

\smallskip
 \textbf{\emph{Case 2}}: The reader receives a message $(sid', \alpha)$ when ${s^j_R} \neq \emptyset$. ${s^j_R} \neq \emptyset$ means that $R$ is currently running a session  $sid$   with partial transcripts ${trs^{sid}} = (sid, c_1, \ldots, c_{\mu}) \in {s^j_R}$, where $\mu \in [1, \gamma]$. $\pi_R$ takes $para$, $k_R$, ${s^j_R}$,  $\mathit{DB}^j$, and $(sid', \alpha)$ as inputs, and proceeds as follows:
 \begin{itemize}
 \item[] (1) If $sid' = sid$, $\mu \in [1, \gamma)$,  and $\alpha$ is valid, then $R$ generates ${s_R^j}'$ including a reply $c_{\mu+1}$, updates ${s_R^j}$ to ${s_R^j}'$, and then sends  $(sid, c_{\mu+1})$ to $T_i$.
 \vspace{1mm}
 \item[] (2) If $sid' = sid$, $\mu = \gamma$, and $\alpha$ is valid, then  $R$ first generates ${s_R^j}'$ including the final $c_{\gamma+1}$, updates ${s_R^j}$ to ${s_R^j}'$, and sends  $(sid, c_{\gamma+1})$ to $T_i$.  Then  $R$ outputs  $o_R=1$, updates $\mathit{DB}^{j}$ to $\mathit{DB}^{j+1}$, and initializes $s_R^{j+1}$ as $\emptyset$.
  \vspace{1mm}
 \item[] (3) If $sid' = sid$,  $\mu \in [1, \gamma]$, and $\alpha$ is invalid, then $R$  outputs  $o_R=0$, updates $\mathit{DB}^{j}$ to $\mathit{DB}^{j+1}$, and initializes $s_R^{j+1}$ as $\emptyset$.
 \vspace{1mm}
 \item[] (4) If $sid' \neq sid$, $R$ simply ignores the message.
 \end{itemize}

 \smallskip
 \textbf{\emph{Case 3}}:
 If $R$ is running a session $sid$ with partial transcripts ${trs^{sid}} = (sid, c_1, \ldots, c_{\mu}) \in {s^j_R}$, where $\mu \in [1, \gamma)$, but fails to receive a message $(sid, \alpha)$ within a pre-fixed time period, then it outputs    $o_R=0$,  updates $\mathit{DB}^{j}$ to $\mathit{DB}^{j+1}$, and initializes $s_R^{j+1}$ as $\emptyset$.

 \smallskip
 \textbf{(2) On tag side: }  Suppose the tag $T_i$ is with current secret key $k_{T_i}^v$ and  internal state $s_{T_i}^v$ for $1 \leq v \leq s$, and there is a coming message $(sid, c)$. $T_i$ deals with $(sid, c)$ as follows.

 \textbf{\emph{Case 1}}: If $c \in P_{c_1}$ (it means that $c$ is a message for starting a new session), there are two subcases.
  \begin{itemize}
  \item[] (1) If $s_{T_i}^v = \emptyset$, then $T_i$ generates  ${s_{T_i}^{v}}'$ including partial transcripts  ${trs^{sid}} = (sid, c, \alpha_1)$ and the random coins for the session $sid$ (if any), where $\alpha_1$ is the reply for $c$. Then $T_i$ updates $s_{T_i}^{v}$ to ${s_{T_i}^{v}}'$, and sends $(sid, \alpha_1)$ to the reader $R$.
  \item[] (2) If $s_{T_i}^v \neq \emptyset$ (it means that $T_i$ is currently running a session),  $T_i$ outputs $o_T  = 0$. Then  $T_i$  erases its internal state $s_{T_i}^{v}$, updates its key to $k_{T_i}^{v+1}$, generates $s_{T_i}^{v+1}$ including  partial transcripts  ${trs^{sid}} = (sid, c, \alpha_1)$ and random coins (if any), where $\alpha_1$ is the reply for $c$.  Finally, $T_i$  sends  $(sid, \alpha_1)$ to the reader $R$.
  \end{itemize}

\textbf{\emph{Case 2}}: $c \notin P_{c_1}$ and $\exists trs^{sid} = (sid, c_1, \ldots, \alpha_\mu) \in s_{T_i}^v$;  there are two sub-cases.
 \begin{itemize}
  \item[] (1) $\mu \in [2, \gamma]$: If $c$ is valid, $T_i$ generates ${s_{T_i}^v}'$ including a reply $\alpha_{\mu}$, updates  $s_{T_i}^v$ with  ${s_{T_i}^v}'$, and sends $(sid, \alpha_{\mu})$ to the reader $R$; else,  $T_i$ outputs $o_{T_i}=0$.  Then $T_i$ erases its internal state $s_{T_i}^{v}$, updates its secret-key $k_{T_i}^{v}$ to $k_{T_i}^{v+1}$, and sets  internal state $s_{T_i}^{v+1} =\emptyset$ for its next session.
  \item[] (2) $\mu = \gamma+1$: If $c$ is valid, $T_i$ sets $o_{T_i}=1$; else, sets $o_{T_i}=0$. Then $T_i$ outputs $o_{T_i}$, erases its internal state $s_{T_i}^{v}$, updates $k_{T_i}^{v}$  to $k_{T_i}^{v+1}$, and sets  $s_{T_i}^{v+1} =\emptyset$.
 \end{itemize}

\textbf{\emph{Case 3}}: In any other cases, $T_i$ simply ignores this message.

 \end{itemize}

\smallskip

Based on the refined model $\mathit{RS}$, we define an RFID system model $RS^*$ that supports tag/mutual authentication and PoP.

\smallskip

\begin{defn}
 An RFID system $RS^*$ that supports  tag/mutual authentication and PoP is defined to be a tuple $(R, \Tt, \func{Setup}^*, \pi^*,$ $\func{CredGen}$, $\func{CredVeri})$.
\end{defn}

\begin{itemize}

 \item $\func{Setup}^*(\kappa,l)$ is defined the same as in Definition \ref{def:rs}.
 \item $\func{Protocol}$ $\pi^*(R, T_i)$ is also defined the same as in Definition \ref{def:rs}.

 \item $\func{CredGen}(para, k_R, s_R^j, DB^j)$: Given inputs $para$, $k_R$, $s_R^j$, and $\mathit{DB}^j$, $\func{CredGen}$ outputs a PoP credential $cred$ if the reader's $j$-th session's execution result $o_R = 1$, else outputs $\bot$.
\smallskip
 \item $\func{CredVeri}(para, cred)$: Given a PoP credential $cred$ and the system's public parameters $para$, it outputs the verification result ``1'' if $cred$ is a valid one, and ``0'' otherwise.

\end{itemize}

\subsection{Adversary model} \label{sec:ouradvmodel}

 A probabilistic polynomial-time concurrent man-in-the-middle (CMIM) adversary $\Aa$
against $RS^*$ can access five oracles, $\func{InitReader}$, $\func{SendT}$, $\func{SendR}$, $\func{CorruptTag}$, and $\func{GetCred}$. The first four oracles are defined the same as in \cite{Deng10}. The oracle $\func{GetCred}$ is for capturing the adversary's capabilities to gain PoP credentials. The adversary can also obtain protocol execution results through $\func{SendT}$ and $\func{SendR}$.
 \smallskip
\begin{itemize}
 \item $\func{InitReader}( )$: $\Aa$ invokes the reader $R$ to start a new session of protocol $\pi^*$, and returns $(sid, c_1)$ to the adversary.

 \item $\func{SendT}(T_i, \hat{sid}, \hat{c})$: $\Aa$ sends a message $(\hat{sid}, \hat{c})$ to the tag $T_i$. $T_i$ returns a message $(\hat{sid}, \alpha)$ (if any) and an execution result $o_{T_i}$ (if any) to the adversary.

\item $\func{SendR}( \hat{sid}, \hat{\alpha})$: $\Aa$ sends a message $(\hat{sid}, \hat{\alpha})$ to the reader. The reader  returns a message $(\hat{sid}, c)$ (if any) and an execution result $o_{R}$ (if any) to the adversary.

\item
$\func{CorruptTag}(T_i)$: $\Aa$ gets all the contents stored on $T_i$, including $T_i$'s current secret key $k_{T_i}$ and internal state $s_{T_i}$.

\item
$\func{GetCred}(sid)$: Given a session ID $sid$, if the reader $R$ has ever been involved in a session with $sid$ as its $j$-th session (without loss of generality), the oracle returns the result of $\func{CredGen}(para, k_R, s_R^j, DB^j)$, or else, returns $\bot$.

\end{itemize}
%$\func{InitReader}$, $\func{SendT}$, $\func{SendR}$, $\func{CorruptTag}$, and $\func{GetCred}$
Let $O_1'$, $O_2'$, $O_3'$, $O_4'$, and $O_5'$ denote the above oracles respectively. And $\Oo'$ denotes the set of them.  An adversary is a
$(t, n_1, n_2, n_3, n_4, n_5)$-adversary, if it makes oracle queries to $O_i'$ without
exceeding $n_i$ times respectively, where $1 \leq i \leq 5$, and the running time is at most $t$.

\subsection{Our new privacy and security notions} \label{section:secNotion}

 Our framework utilizes two existing security notions defined in \cite{Deng10}, namely  adaptive completeness and tag/mutual authentication, and  two new notions,  unp$^\#$-privacy and PoP credential unforgeability.

\smallskip

\subsubsection{Unp$^\#$-privacy}

Intuitively, unp$^\#$-privacy requires that: (1) protocol messages should be pseudorandom; (2) every message (except the first one which normally contains a random value) should be authenticated. Compared with unp$^\tau$-privacy, it is not constrained to analyse three-round protocols only, it is defined under a more reasonable adversary model, and it is defined with delicate session control when $b=0$ holds in the guess stage.

 The unp$^\#$-privacy experiment ${\bf Exp}_{\Aa}^{\mathit{unp}^\#}[\kappa, n_1, n_2, n_3, n_4,$ $n_5]$ is illustrated in Figure \ref{Figure:unpsharp}. To conduct delicate session control, we define two types of sessions. If a session $sid$ is initialized through querying $O_1'$, it is considered as a session between the reader  $R$  and the target tag  $T_c$. Its transcripts are recorded with $trs^{sid}$. Later, when the adversary sends a message $(sid, c)$ to $T_c$ through $O_2'$, the experiment checks $trs^{sid}$: if the message $c$ is the same as the last message $c_{\mu}$ stored in $trs^{sid}$, which means that $c$ came from the reader $R$ without being changed, the experiment returns the next message $\alpha_{\mu} \in_R P_{\alpha_{\mu}}$   (if $ 1 \leq \mu \leq \gamma$) or $o_T = 1$ (if $\mu = \gamma+1$), and updates $trs^{sid}$ by adding $\alpha_{\mu}$ or $o_T$ ; or else, the message $c$ is considered invalid, the experiment returns $o_T = 0$, and updates $trs^{sid}$ by adding $o_T$. The queries to $O_3'$ are processed similarly.
The second type of sessions are the ones initialized through querying $O_2'$. An adversary chooses $(sid, c)$ and queries $O_2'$ to start a session with $T_c$. The transcripts are recorded with $trsAdv^{sid}$. The adversary is allowed to get a response from $O_2'$, but if it queries $O_2'$ again with the third round protocol message $(sid, c')$, the experiment would output $o_T = 0$ and terminate the session.

\newsavebox{\unpsharp}
\begin{lrbox}{\unpsharp}
\begin{tabular}{|l|}
\hline\\[-0.8em]
 Experiment {\bf Exp$_{\Aa}^{\mathit{unp}^\#}[\kappa, n_1, n_2, n_3, n_4, n_5]$}\\
 1. run \textsf{Setup($\kappa$)} to setup $(R,\Tt)$.\\
 ~~//learning stage \\
 2. $\{T_c, st\} \leftarrow \Aa_1^{\Oo'}(R, \Tt)$. \\
 3. select $b \in_R \{0, 1\}$.  \\
 ~~//guess stage\\
 4. $b' \leftarrow \Aa_2\{O_1',O_2',O_3'\}(R, T_c, st)$; \\
 ~(1) if $b = 1$,  \\
 ~ when $\Aa_2$ queries $O_1'$,$O_2'$, and $O_3'$, return the results of\\
 ~ the oracles. \\
 ~(2) if $b = 0$, \\
 ~a. when $\Aa_2$ queries $O_1'$, return $sid \in_R P_{sid}$ and\\
 ~~~~ $c_1 \in_R P_{c_1}$, create $trs^{sid} = (sid, c_1)$.\\
 ~b. when $\Aa_2$ queries $O_2'$ with parameters $(sid, c)$, \\
 ~~~~if $\nexists trs^{sid} \wedge \nexists trsAdv^{sid} \wedge c \in P_{c_1}$, then \\
 ~~~~~~choose $\alpha \in_R P_{\alpha_1}$, create $trsAdv^{sid} = (sid, c,\alpha)$,\\
 ~~~~~~return $(sid, \alpha)$; \\
 ~~~~if $\exists trs^{sid} = (sid, \ldots, c_\mu) \wedge \mu < \gamma+1$, then \\
 ~~~~~~if $c_\mu = c$, then choose $\alpha \in_R P_{\alpha_{\mu}}$, \\
 ~~~~~~~~~update $trs^{sid} = trs^{sid}\uplus\alpha $, return $(sid, \alpha)$;\\
 ~~~~~($trs^{sid}\uplus\alpha$ means that $\alpha$ is appended to $trs^{sid}$) \\
 ~~~~~~else\\ %$c_\mu \neq c$
 ~~~~~~~~~if $\mu = 1 \wedge c \in P_{c_1}$, \\
 ~~~~~~~~~//$c_1$ from $R$ has been modified to $c$, then the ``tag''\\
 ~~~~~~~~~//responds with $\alpha$. $\alpha$ will be rejected by the\\
 ~~~~~~~~~//reader later, so $o_R$ is set as 0 in advance.\\
 ~~~~~~~~~~~then choose $\alpha \in_R P_{\alpha_{1}}$, set $o_R =0$, \\
 ~~~~~~~~~~~update $trs^{sid} = trs^{sid}\uplus \alpha \uplus o_R$, \\
 ~~~~~~~~~~~return $(sid, \alpha)$;\\
 ~~~~~~~~~else set $o_T = 0$, update $trs^{sid} = trs^{sid} \uplus o_T$,\\ % two cases£º
 ~~~~~~~~~~~return $o_T$;\\
 ~~~~if $\exists trs^{sid} = (sid, \ldots, c_{\gamma+1}, o_R=1)$, then \\
 ~~~~~~if $c_{\gamma+1} = c$, then set $o_T=1$; else, set $o_T=0$;\\
 ~~~~~~update $trs^{sid} = trs^{sid} \uplus o_T$, return $o_T$;\\
 ~~~~if $\exists trsAdv^{sid} = (sid, c_1, \alpha_1)$, then \\
 ~~~~~~set $o_T=0$, \\
 ~~~~~~update $trsAdv^{sid} = trsAdv^{sid}\uplus c\uplus o_T$,\\
 ~~~~~~return $o_T$;\\
 ~~~~else ignore this query;\\
 ~c. when $\Aa_2$ queries $O_3'$ with parameters $(sid, \alpha)$, \\
 ~~~~if $\nexists trs^{sid}$, then ignore this query;\\
 ~~~~if $\exists trs^{sid} = (sid, c_1, \alpha_1, o_R=0)$, then\\
 ~~~~~~return $o_R = 0$;\\
 ~~~~if $\exists trs^{sid} = (sid, c_1, \ldots, \alpha_\mu)$, then\\
 ~~~~~~~if $\alpha = \alpha_\mu$, then choose $c \in_R P_{c_{\mu+1}}$,\\
 ~~~~~~~~~if $\mu < \gamma$, then update $trs^{sid} = trs^{sid}\uplus c, $\\
 ~~~~~~~~~~~~return $(sid, c)$;\\
 ~~~~~~~~~else \\
 ~~~~~~~~~~~~set $o_R = 1$, \\
 ~~~~~~~~~~~~update $trs^{sid} = trs^{sid}\uplus c\uplus o_R$,\\
 ~~~~~~~~~~~~return $(sid, c)$ and $o_R$;\\
 ~~~~~~~else set $o_R = 0$, \\
 ~~~~~~~~~~~~update $trs^{sid} = trs^{sid}\uplus o_R$,\\
 ~~~~~~~~~~~~return $o_R$;\\

 ~~~~else ignore this query. \\
 5. output 1 if $b' = b$, and 0 otherwise.\\
\hline
\end{tabular}
\end{lrbox}
\begin{figure}[htbp]
 \centering  \scalebox{0.9}
{\usebox{\unpsharp}} \caption{Unp$^\#$-Privacy Experiment
}\label{Figure:unpsharp}
\end{figure}

\begin{defn}
The advantage of adversary $\Aa$ in the experiment $\mathbf{Exp}^{unp^\#}_\Aa$
 is defined as:
\begin{eqnarray*}
% \nonumber to remove numbering (before each equation)
&&Adv^{unp^\#}
_\Aa(\kappa, l, n_1, n_2, n_3, n_4, n_5) \\
 &=&|\Pr[\mathbf{Exp}^{unp^\#}_\Aa(\kappa, l, n_1, n_2, n_3, n_4, n_5) = 1] - \frac{1}{2}|,
\end{eqnarray*}
where the probability is taken over the choice of the tag set $\Tt$ and the coin tosses of the adversary
$\Aa$.
\end{defn}

\smallskip
\begin{defn}
An adversary $\Aa$ $(\epsilon, t, n_1, n_2, n_3, n_4, n_5)$-breaks the unp$^\#$-privacy of the RFID system
$(R, \Tt)$ if the advantage $Adv^{unp^\#}(\kappa, l, n_1, n_2, n_3, n_4, n_5)$ of $\Aa$ in the experiment $\mathbf{Exp}^{unp^\#}_\Aa$ is at least $\epsilon$,
 and the running time of $\Aa$ is at most $t$.
\end{defn}

\smallskip
\begin{defn}
[Unp$^\#$-Privacy] An RFID system $(R,\Tt)$ is said to be $(\epsilon, t, n_1, n_2, n_3, n_4, n_5)$-unp$^\#$-private,  if  for all sufficiently large $\kappa$, there exists no adversary who can $(\epsilon, t, n_1, n_2, n_3, n_4 , n_5)$-break the unp$^\#$-privacy of $(R,\Tt)$ for any $(\epsilon, t)$, where $t$ is polynomial in $\kappa$ and $\epsilon$ is non-negligible in $\kappa$.
\end{defn}

\subsubsection{PoP credential unforgeability}
Intuitively, PoP credential unforgeability has two folds of meanings. Each credential regarding the reader and a valid tag should correspond to a completed protocol session between the reader and the tag. And an adversary should not be able to generate a valid PoP credential regarding itself and a valid tag.

PoP credential unforgeability is formalized by the experiment
${\bf Exp}_{\Aa}^{\mathit{CredUfrg}}[\kappa,l]$ shown in Figure \ref{fig:CUfrg}. The adversary
$\Aa$ is allowed to query the five oracles without exceeding
$n_1,n_2,n_3,n_4, n_5$ times respectively. Then it outputs a PoP credential $cred$. Denote by $E_1$ the event under both of the following conditions: (1) the PoP credential corresponds to the reader $R$ and an uncorrupted tag $T_i$; (2) for the reader $R$, there does not exist $j \in [1, ls]$,  such that $\func{CredGen}(para, k_R, s_R^j, DB^j)=cred$ and $\func{CredVeri}(para, cred) = 1$. %\blue{This requires that a valid reader should have a matching session with each tag for generating a valid credential.}
Denote by $E_2$ the event under the condition that the PoP credential corresponds to an adversary  $\Aa$ and an uncorrupted tag $T_i$  and $\func{CredVeri}(para\cup papa_\Aa, cred) = 1$,  where $para_\Aa$ are the adversary $\Aa$'s public parameters.

\smallskip

\newsavebox{\CUfrg}
\begin{lrbox}{\CUfrg}
\begin{tabular}{|l|}
\hline\\[-0.8em]

 Experiment {${\bf Exp}_{\Aa}^{\mathit{CredUfrg}}[\kappa,l,n_1,n_2,n_3,n_4, n_5]$}\\
 1. run \textsf{Setup($\kappa, l$)} to setup $(R, \Tt)$.  \\
 2. $cred \leftarrow \Aa^{\Oo'}(R, \Tt)$.\\ %//
\hline
\end{tabular}
\end{lrbox}

\begin{figure} [htbp]
\centering \scalebox{0.9}
{\usebox{\CUfrg}} \caption{PoP Credential Unforgeability Experiment}\label{fig:CUfrg}
\end{figure}

\begin{defn}
An adversary $\Aa$ $(\epsilon, t, n_1, n_2, n_3, n_4, n_5)$-breaks PoP credential unforgeability of an RFID system $(R, \Tt)$ if the probability that event $E_1$ or $E_2$ occurs is
at least $\epsilon$ with the running time at most $t$, and $\Aa$ is a $(\kappa, l, n_1, n_2, n_3, n_4, n_5)$-adversary.
\end{defn}

\smallskip
\begin{defn}
[Credential unforgeability]
The RFID system $(R, \Tt)$ satisfies PoP credential unforgeability, if for all sufficiently large $\kappa$ there exists no adversary $\Aa$ that can $(\epsilon, t, n_1, n_2, n_3, n_4, n_5)$-break the credential unforgeability of $(R, \Tt)$ for any $(\epsilon, t)$, where $t$ is polynomial in $\kappa$ and $\epsilon$ is non-negligible in $\kappa$.
\end{defn}

%!TEX root = main.tex
\section{Generic construction}
In this section, we provide a generic construction to transform an RFID tag/mutual authentication protocol to one with PoP, conduct security analysis, and discuss its practicability.

\vspace{-2mm}
\subsection{Cryptographic primitives}

The cryptographic primitives used in our construction include  pseudorandom function, cryptographic hash function, and digital signature scheme.

\smallskip
\textbf{Pseudorandom function (PRF)} \cite{Li2011}
%Let $F : \Kk \times \Dd \rightarrow \Rr$ be a family of functions, where $\Kk$
%is the set of keys (or indexes) of $F$, $\Dd$ is the domain of
%$F$, and $\Rr$ is the range of $F$. Let $\mathsf{Rand}^{\Dd \rightarrow \Rr}$ be the family of all functions with domain $\Dd$ and range $\Rr$. Let $|\Kk| = \gamma$, $|\Dd| = m$, and
%$|R| = n$. A \emph{polynomial time predictable test} (PTPT) for
%$F$ is an experiment, where a probabilistic polynomial time
%algorithm $T$, given $\gamma$, $m$, $n$, $j$ as inputs and with access to
%an oracle $\Oo_f$ for a function $f \in F$, outputs either 0 or 1. Figure \ref{Figure:prf} shows a \emph{PTPT} for $F$. At first, algorithm $T$ queries
%the oracle $\Oo_f$ about $x_1, \ldots, x_j$. Then, it outputs $x \in \Dd$ such
%that $x \neq x_1, \ldots, x_j$. This $x$ is called the chosen exam. At
%this point, algorithm $T$ is not allowed to query oracle $\Oo_f$
%any more. The experiment chooses  $b \in_R \{0, 1\}$.
%If $b = 1$, then $y = f(x)$; otherwise,
%$y \in_R \Rr$. Then $y$ is given to $T$. Finally,   algorithm $T$ is required to
%output a bit $b'$ by guessing which of the two values is given
%to it: $b' = 1$ for $y = f(x)$, and $b' = 0$ for $y \in_R \Rr$.
 Let $F : \Kk \times \Dd \rightarrow \Rr$ be a family of functions, where $\Kk$, $\Dd$, and $\Rr$ denote the set of keys (or indexes), the domain, and
 the range of $F$ respectively. Let $|\Kk| = \gamma$, $|\Dd| = m$, and
$|R| = n$. Let $\mathsf{Rand}^{\Dd \rightarrow \Rr}$ be the family of all functions with domain $\Dd$ and range $\Rr$. A \emph{polynomial time predictable test} (\emph{PTPT}) for $F$ is an experiment, where  a probabilistic polynomial time algorithm $T$, given $\gamma$, $m$, $n$ as input  and with access to an oracle $O_f$ for a function $f \in_R F$ or $f \in_R \mathsf{Rand}^{\Dd \rightarrow \Rr}$, outputs either 0 or 1.
 Figure \ref{Figure:prf} shows a \emph{PTPT} for $F$.

\newsavebox{\prf}
\begin{lrbox}{\prf}
\begin{tabular}{|l|}
\hline\\[-0.8em]
 Experiment {\bf Exp$_{T}^{ptpt}(F,\gamma, m,n)$}\\
 1. select $b\in_R \{0,1\}$.  \\
 2. if $b=1$,  select $k \in_R \Kk$ and set $f=F_k$; \\
 ~~~otherwise, $f \in_R  \mathsf{Rand}^{\Dd \rightarrow \Rr}$.  \\
 3. $b' \leftarrow T^{O_f}$.  \\
\hline
\end{tabular}
\end{lrbox}

\begin{figure}[htbp]
\centering \scalebox{0.9}
{\usebox{\prf}} \caption{Polynomial Time Predictable Test
}\label{Figure:prf}
\end{figure}

\smallskip
\begin{defn} An algorithm $T$ passes the \emph{PTPT} for the
function family $F$ if it correctly guesses the random bit which is selected by the \emph{PTPT} experiment, namely $b' = b$.
The advantage of algorithm $T$ is defined as
\begin{equation*}
 Adv_T (\gamma, m, n) = |Pr[b' = b]-\frac{1}{2}|,
\end{equation*}
where the probability is taken over the choice of $f$ in $F$ and
the coin tosses of algorithm $T$.
\end{defn}

\newpage
\begin{defn} A function family $F : \Kk \times \Dd \rightarrow \Rr$ is said
to be a pseudorandom function family if it has the following
properties:
\begin{itemize}
 \item \emph{Indexing}: Each function in $F$ has a unique $\gamma$-bit key (index)
associated with it. It is easy to select a function $f \in F$
randomly if $\gamma$ random bits are available.
 \item \emph{Polynomial Time Evaluation}: There exists a polynomial time
algorithm such that, given input of a key (index) $k \in \Kk$
and an argument $x \in \Dd$, it outputs $F_k(x)$.
 \item \emph{Pseudorandomness}:
No probabilistic polynomial time algorithm $T$ can pass the \emph{PTPT} for $F$ with non-negligible
advantage.
 \end{itemize}
\end{defn}

\smallskip
 \textbf{Digital signature scheme} A digital signature scheme $\mathcal{DS} = (\Kk, \Ss, \Vv)$ consists of three algorithms: (1) a randomized key generation algorithm $\mathcal{K}$: $(\mathit{PK},sk) \stackrel{\$}{\leftarrow} \mathcal{K}(\kappa)$
which returns a pair of  public key and private key  $(\mathit{PK},sk)$ given a security parameter $\kappa$; (2) a signing algorithm $\mathcal{S}$: $\sigma
 \stackrel{\$}{\leftarrow} \mathcal{S}_{sk}(M)$ (may be randomized or stateful) that takes a secret key $sk$ and a message $M$ as input   and outputs a signature $\sigma \in
\{0, 1\}^* \cup \{\bot\}$; (3) a deterministic verification algorithm $\Vv$: $d \leftarrow \Vv_{\mathit{PK}}(M, \sigma)$ that takes a public key $\mathit{PK}$, a message $M$, and a signature $\sigma$ as input  and outputs a bit $d$, where $d = 1$ if $\sigma$ is valid and $d = 0$ if $\sigma$ is invalid. Existential unforgeability under an adaptive chosen-message attack (EU-CMA) \cite{Goldwasser88} is a widely used security notion for digital signature scheme.

\subsection{Generic construction} \label{sec:genConstr}

%Suppose there is an RFID system $RS = (R, \Tt, \func{Setup}, \pi)$. $RS$ contains a reader $R$ and a set of $l$ tags $\Tt$. $\pi(R, T_i)$ is a symmetric-key based tag/mutual authentication protocol that satisfies adaptive completeness, tag/mutual authentication, and unp$^\#$-privacy. We can transform $RS = (R, \Tt, \func{Setup}, \pi)$ (see Definition \ref{def:rs}) to a new system $RS^*=(R, \Tt, \func{Setup^*}, \pi^*,$ $\func{CredGen}, \func{CredVeri})$, where $\pi^*$ supports PoP and satisfies adaptive completeness, tag/mutual authentication, PoP credential unforgeability, and unp$^\#$-privacy. The transforming procedures are defined as follows.

Given an RFID system  $RS = (R, \Tt, \func{Setup}, \pi)$ as defined in Definition \ref{def:rs}, our construction $RS^*=(R, \Tt, \func{Setup^*}, \pi^*,$ $\func{CredGen}, \func{CredVeri})$ is shown below. % with symmetric-key based tag/mutual authentication, we construct a new system with mutual authentication and PoP as follows.

 \paragraph{$\func{Setup} \rightarrow \func{Setup^*}$} $RS^*$ introduces a signature scheme $DS$, a cryptographic hash function $H: \{0,1\}^{*} \longrightarrow \{0,1\}^{l_g}$, and a PRF family $G: \{0,1\}^{l_k} \times \{0,1\}^{2l_g} \longrightarrow \{0,1\}^{2l_g}$ as additional building blocks, where $l_k$ and $l_g$ are polynomial to the security parameter $\kappa$ of $RS$.
 Each tag is embedded with an additional  $l_k$-bit key $k'_{T_i}$ with an initial value $k^{'1}_{T_i}$. Each tag and the reader are assigned with
 a unique pair of public key and secret key for the signature scheme, denoted as $(\mathit{PK}_{T_i}, sk_{T_i})$ for $1 \leq i \leq l$, and $(\mathit{PK}_{R}, sk_{R})$ respectively.
 The reader $R$'s secret key, initial state, and each tag $T_i$'s secret key, initial state, public parameters are updated/copied from $RS$'s settings accordingly: $k^*_{R} = \{k_{R}, sk_R\}$, ${s_R^1}^* = s_R^1$, ${k_{T_i}^1}^* = \{k_{T_i}^1, k^{'1}_{T_i}, sk_{T_i}\}$, ${s_{T_i}^1}^* = s_{T_i}^1$, $\xi_{T_i}^* = \{\xi_{T_i}, \mathit{PK}_{T_i}\}$.
 ${DB^1}^*$ is updated such that for each $T_i$, $rcd_{T_i}^* = \{rcd_{T_i}, k^{'1}_{T_i}, \mathit{PK}_{T_i}\}$, where $rcd_{T_i} \in DB^1$.
 $RS^*$'s public system parameters $para^*$ are set as $\{\sigma^*, \mathit{PK}_{R}, \xi_{T_1}^*, \ldots, \xi_{T_l}^*\}$, where $\sigma^*$ consists of  $\sigma$, and all public parameters from the three new building blocks, $DS$, $H$, and $G$.

 \paragraph{$\pi \rightarrow \pi^*$} Given $\pi(R, T_i) = \{c_1, \ldots, \alpha_{\gamma}, c_{\gamma+1},$ $o_R, o_{T_i}\}$, $\pi^*(R, T_i)$ is constructed as $\{c_1, \cdots, c_\gamma, \alpha_\gamma$, $c_{\gamma+1}^*$, $\alpha_{\gamma+1}^*$, $o_R^*,$ $o_{T_i}^*\}$. The superscript $\star$ denotes that the procedures for generating the message/result are different from $\pi$ or  newly added. Note that $c_{\gamma+1}$ could be null if $\pi$ consists of only $2\gamma$ rounds. And in the protocol $\pi$, the reader $R$ outputs $o_R$ and updates $T_i$'s secret key after sending $c_{\gamma+1}$, while in the protocol $\pi^*$, the reader $R$ postpones these two steps and any other updates that depend on the reader's execution result $o_R$ till receiving $\alpha_{\gamma+1}^*$. Similarly, the tag $T_i$ also postpones outputting $o_{T_i}$ and updating $T_i$'s secret key till it sends the last message $\alpha_{\gamma+1}^*$.

\begin{figure}[!htbp]
 \centering
 \includegraphics[scale=0.9]{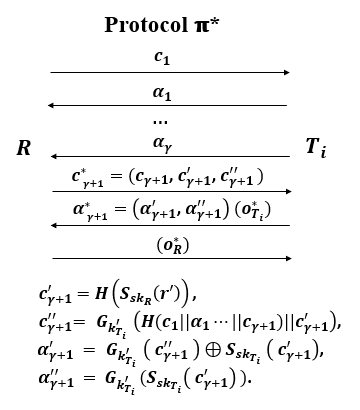}\\
 \caption{An RFID Protocol with PoP}\label{fig:gp2}
\end{figure}

 The protocol $\pi^*$ is shown in Fig. \ref{fig:gp2}.   We omit $sid$ in each message, and put $o^*_R$/$o^*_{T_i}$ in parentheses to demonstrate that it is sent at the same time with the corresponding message (if any) but may through a different channel.  The protocol $\pi^*$'s procedures are the same as the protocol $\pi$ in generating and dealing with the first $2\gamma-1$ messages. Suppose that $R$ currently is with secret key $sk_R$, internal state $s_R$ and database $\mathit{DB}$, and $T_i$ is with secret keys  $k_{T_i}$, $k'_{T_i}$ and $sk_{T_i}$, and internal state $s_{T_i}$.
Below we show the  procedure  dealing with the $2\gamma$-th round message $\alpha_\gamma$ and onwards.

 \begin{itemize}

 \item Upon receiving $\alpha_\gamma$ ($sid$ is omitted for simplicity), the reader $R$  authenticates $T_i$ with $\alpha_\gamma$ and  computes $c_{\gamma+1}$ according to $\pi$'s procedures first. Then $R$  computes $c_{\gamma+1}'= H(\Ss_{sk_R}(r))$, $c_{\gamma+1}''= G_{k'_{T_i}}(H(c_1||\alpha_1\cdots||c_{\gamma+1})||c_{\gamma+1}')$. %, where $\Ss_{sk_R}(r)$ is the reader's signature on $r$.
     For simplicity, we set $r \in_R \{0, 1\}^{l_g}$ here. In real applications, $r$ could be a string with arbitrary length, and contains detailed information about the session, for example, an authenticated timestamp.
 Finally, $R$ sends $c_{\gamma+1}^*=(c_{\gamma+1}, c_{\gamma+1}', c_{\gamma+1}'')$ to $T_i$.

\item Upon receiving $c_{\gamma+1}^* = (c _{\gamma+1}, c_{\gamma+1}', c_{\gamma+1}'')$, if $c_{\gamma+1}$ is valid according to the protocol $\pi$'s procedures and $c_{\gamma+1}'' = G_{k'_{T_i}}(H(c_1||\alpha_1\cdots||c_{\gamma+1})||c_{\gamma+1}')$, $T_i$ sets $o_{T_i}^* = 1$; else sets $o_{T_i}^* = 0$. If $o_{T_i}^* = 1$, $T_i$ computes $\alpha_{\gamma+1}' = G_{k'_{T_i}}(c_{\gamma+1}'') \oplus \Ss_{sk_{T_i}}(c_{\gamma+1}')$ and $\alpha_{\gamma+1}'' = G_{k'_{T_i}}(\Ss_{sk_{T_i}}(c_{\gamma+1}'))$,  then sends $\alpha_{\gamma+1}^*= (\alpha_{\gamma+1}',$ $\alpha_{\gamma+1}'')$ to  $R$. Finally,  $T_i$  outputs  $o_{T_i}^*$, erases $s_{T_i}$, updates $k_{T_i}$ according to the procedures in $\pi$, and initializes internal state for the next session as $\emptyset$. %$k_{T_i}'$ will not be updated in this construction.

 \item Upon receiving $\alpha_{\gamma+1}^* = (\alpha_{\gamma+1}', \alpha_{\gamma+1}'')$,  $R$ computes $\sigma = \alpha_{\gamma+1}'\oplus G_{k'_{T_i}}(c_{\gamma+1}'')$. Then $R$ sets  $o_R^* =1$ if $\Vv_{\mathit{PK}_{T_i}}(c_{\gamma+1}', \sigma) = 1$ and $\alpha_{\gamma+1}'' = G_{k'_{T_i}}(\sigma)$; else,  sets $o_R^* =0$. And finally, $R$ outputs $o_R^*$, and updates its internal state and database.

 \end{itemize}

 \paragraph{$\func{CredGen}(para^*, {k_R}^*, {s_R^j}^*, {DB^j}^*)$} Given system parameters $para^*$, the reader's secret key ${k_R}^*$, and the $j$-th session's internal state ${s_R^j}^*$ and database ${DB^j}^*$, the function outputs a PoP credential $cred$ if the reader has completed the $j$-th session with $o_R^* = 1$, where $o_R^* \in {s_R^j}^*$; or else, outputs ``$\bot$''. ${s_R^j}^*$ contain the $j$-th session's transcripts  $trs = \{c_1, \ldots, \alpha_\gamma$, $c_{\gamma+1}^*=(c _{\gamma+1},c_{\gamma+1}',c_{\gamma+1}'')$, $\alpha_{\gamma+1}^* = (\alpha_{\gamma+1}', \alpha_{\gamma+1}'')$, $o_R^*$, $o_{T_i}^*\}$  and random coins which include $r'$ for generating $c_{\gamma+1}'$.
 Suppose $R$ has authenticated the tag as $T_i$, $cred$ is computed as a tuple $\{R, T_i, cred_1, cred_2, cred_3\}$, where $cred_1 = r'$, $cred_2 = \Ss_{sk_R}(r')$, and $cred_3 = \alpha_{\gamma+1}'\oplus G_{k'_{T_i}}(c_{\gamma+1}'')$.

 \paragraph{$\func{CredVeri}(para^*, cred)$} Given a PoP credential $cred = \{R, T_i, cred_1, cred_2, cred_3\}$ and the system's public parameters $para^*$, it outputs the verification result ``1'' if $\Vv_{\mathit{PK}_R}(cred_1, cred_2) \wedge \Vv_{\mathit{PK}_{T_i}}(H(cred_2), cred_3) = 1$, and ``0'' otherwise, where $\mathit{PK}_R$ and $\mathit{PK}_{T_i}$ are retrieved from $para^*$.

 \subsection{Security analysis on the generic construction}
$\pi^*(R, T_i)$ can be considered as a combination of two  protocols, including the underlying protocol $\pi(R, T_i)$ and a new protocol $\pi'(R, T_i)$.
 $\pi'(R, T_i)$ is a two-round mutual authentication protocol. $R$ and $T_i$'s share a secret key $k'_{T_i} \in_R \{0,1\}^{l_k}$ and a string $trs \in_R \{0,1\}^{l_\pi}$, where $l_\pi$ is the length of $\pi$'s transcripts. $R$ also has $T_i$'s public key $\mathit{PK}_{T_i}$.
$\pi'$  runs as the following.

\begin{enumerate}
   \item The reader $R$ chooses $r' \in_R \{0,1\}^{l_g}$, computes $c'=H(\Ss_{\mathit{PK}_R}(r'))$ and $c''=G_{k'_{T_i}}(H(trs)||c'))$. Then it sends $c = (c', c'')$ to $T_i$.
   \item  Upon receiving $c = (c', c'')$, if $c''=G_{k'_{T_i}}(H(trs)||c'))$, $T_i$ computes $\alpha' = G_{k'_{T_j}}(c'')\oplus \Ss_{sk_{T_i}}(c')$  and $\alpha'' = G_{k'_{T_j}}(\Ss_{sk_{T_i}}(c'))$, sends $\alpha = (\alpha', \alpha'')$ to $R$, and outputs $o_T' = 1$; else, outputs $o_T' = 0$.
   \item  Upon receiving $\alpha = (\alpha', \alpha'')$, $R$ computes $\sigma = \alpha'\oplus G_{k'_{T_i}}(c'')$ first. Then if $\Vv_{\mathit{PK}_{T_i}}(c', \sigma) = 1$ and $\alpha'' = G_{k'_{T_i}}(\sigma)$, the reader $R$ outputs $o_R' =1$; else, outputs $o_R' =0$.
  \end{enumerate}

\smallskip
In $\pi^*(R, T_i)$, $R$ and $T_i$ first run the protocol $\pi$'s procedures to authenticate $T_i$. After receiving $\alpha_{\gamma}$, if $R$ considers $T_i$ as a valid tag, say $T_j$ ( $T_j =T_i$ if the authentication is correct, else $T_j \neq T_i$), it sets the value of $trs$ as the concatenation of $\pi$'s protocol messages, and continues to run $\pi'$'s procedures with $T_i$   using $T_j$'s keys $k'_{T_j}$ and $\mathit{PK}_{T_j}$.

\smallskip

\smallskip
We conduct security analysis on the protocol $\pi'(R, T_i)$ and the combined protocol  $\pi^*(R, T_i)$, and have the following theorems (proof sketches  are provided in Appendix).

\smallskip
\begin{theorem} \label{theo:pipall}
If  the function family $G:\{0,1\}^{l_k} \times \{0,1\}^{2l_g} \rightarrow \{0,1\}^{2l_g}$ is a PRF family, and the underlying signature is complete, then  $\pi'(R, T_i)$  satisfies completeness,  mutual authentication, and unp$^\#$-privacy under the random oracle model.
\end{theorem}

\smallskip
%\begin{theorem} \label{theo:pistarall}
%If both of  $\pi(R, T_i)$ and   $\pi'(R, T_i)$ satisfy adaptive completeness, mutual authentication, and unp$^\#$-privacy, and  the underlying signature is EU-CMA secure, then  $RS^* = (R, \Tt, \func{Setup}^*, \pi^*,$ $\func{CredGen}$, $\func{CredVeri})$  satisfies adaptive completeness,   mutual authentication, PoP credential unforgeability, and unp$^\#$-privacy.
%\end{theorem}

\begin{theorem} \label{theo:pistarall}
If  $\pi(R, T_i)$ satisfies tag/mutual authentication, $\pi'(R, T_i)$  satisfies mutual authentication,  both of $\pi(R, T_i)$ and $\pi'(R, T_i)$   satisfy adaptive completeness  and unp$^\#$-privacy, and  the underlying signature is   EU-CMA secure, then  $RS^* = (R, \Tt, \func{Setup}^*, \pi^*,$ $\func{CredGen}$, $\func{CredVeri})$  satisfies adaptive completeness,   mutual authentication, PoP credential unforgeability, and unp$^\#$-privacy.
\end{theorem}

\subsection{On the practicability of the generic construction}

Unavoidably, our generic construction incurs additional requirements and runtime overheads.
Among them,  computational requirements and runtime overheads on the tag side are more critical.

In terms of computational capability, our generic construction requires tags to support  hash, PRF, and signature scheme.
 Li et al. \cite{Li2011} pointed out that, to achieve unp$^*$-privacy, RFID tags at least should have  the ability to compute a PRF or its equivalents such as symmetric block ciphers or cryptographic hash functions. This minimum condition applies to unp$^\#$-privacy as well.  A compact hash function with 128-bit output requires about 4,000 gate equivalents \cite{Bogdanov08}, and a compact AES 128-bit implementation requires 3,400 gate equivalents \cite{Feldhofer04}.  There exist  passive RFID tags (e.g., \cite{hid,nxp}) that support AES on the market.

Asymmetric cryptographic primitives  normally are considered too heavy for low-cost RFID tags. Fortunately,  ECC has been demonstrated to be usable in passive UHF RFID systems \cite{Liu15, Tan17}. Compared to other types of signature schemes % (e.g., RSA \cite{Rivest78})
 with the same security level, ECC-based ones enjoy small key sizes and signature sizes, and   better efficiency.  It is feasible to adopt an ECC-based signature scheme in our construction due to the following reasons.
First, efficient ECC processors have been proposed for passive ultra-high frequency (UHF) RFID  tags. For example, Tan et al. \cite{Tan17} designed  a 163-bit ECC processor (with 80-bit security) which requires around 12,000 gate equivalents.  %In the case that the protocol $\pi$ is hash-based and the PRF is instantiated with the same hash function, our construction  requires RFID tags to be equipped with one hash function and one ECC processor,  which totally take less than 16,000 gate equivalents for 80-bit security.
Second, in some ECC-based signature schemes, the expensive  ECC point multiplication operations can be eliminated by pre-computing them.  %For example, for the ECC-based Schnorr signature scheme \cite{Schnorr91} with parameters $(E(\Ff_q),P, t,h)$, where $P$ is a point on the elliptic curve $E(\Ff_q)$ with order $t$,  $\Ff_q$ is a finite field with order $q$, and  $h()$ is a hash function such that $h: \{0,1\}^* \rightarrow \Zz_q$,  the signer  can pre-compute several pairs of $(r, rP)$ and store them for generating signatures, where $r \in_R [1,t]$.  However, as each signature consumes one unused pair of $(r, rP)$, this approach incurs linear  storage to
% the signer.
Third, there exist ECC-based signature schemes specifically designed for resource-constrained devices, such as  Yavuz and Ozmen's $K$-time ECC-based signature scheme SEMECS  \cite{Yavuz19}.% that achieves optimal signature and private key sizes, and requires constant-size storage and no ECC operation at the signer. %However, there are some tradeoffs.  SEMECS only supports generating $K$ signatures at most for each secret key, and the public key's size is linear to $K$.
%Although in our generic  construction, the reader and the tags adopt the same signature scheme, it is also feasible that the reader adopts  a full-time signature  and the tags adopt a $K$-time  one, as long as   the full-time signature scheme satisfies  completeness and EU-CMA,  and the $K$-time signature scheme satisfies  completeness and $K$-time EU-CMA . %(e.g., the ECC-based Schnorr signature scheme)  (e.g., SEMECS)

The modularized design of our generic construction  makes the best use of existing tag/mutual authentication protocols, and  simplifies the security analysis. In practice, it may not be necessary to generate a PoP credential in each and every  run of the protocol. In such case, our construction can be easily implemented to support  two running modes for better efficiency. In one mode,  it only runs the underlying protocol $\pi$'s procedure for tag/mutual authentication  if no PoP credential is required; in the other  mode, it runs  the full procedure if PoP credential is needed.

\section{An RFID system with mutual authentication and PoP} \label{Section:protocol}

In this section, we  refine the RFID mutual authentication protocol proposed by Li et al. in \cite{Li2011}, and prove that the refined protocol is secure and satisfies unp$^\#$-privacy. Then we construct a protocol to support PoP based on the refined protocol, and prove that the constructed protocol is secure and of unp$^\#$-privacy under our framework.

\subsection{A refined RFID mutual authentication protocol}

Compared with the original  RFID mutual authentication protocol proposed by Li et al. \cite{Li2011}, our  refined protocol   is different in two aspects: (1) the reader $R$ does not need to send a dummy third round message if the second round message from the tag is invalid; (2) both the reader and the tag explicitly output an  execution result. The refined system $\mathit{MA} = (R, \Tt, \func{Setup}, \pi)$ is defined below.

\setcounter{paragraph}{0}
\subsubsection{$\func{Setup}(\kappa,l)$} It initializes the whole system $RS$ with a single legitimate reader $R$ and $l$ tags $\Tt$. $l$, $l_k$, $l_d$, $l_{p}$, $l_{v}$, $l_r$, $l_u$ are all polynomial in the security parameter $\kappa$, where  $l_r+l_{p}=l_d$ and  $l_u+l_r+l_{v}=l_d$. $pad$ is a pre-fixed $l_p$-bit padding.
 Let $F: \{0, 1\}^{l_k} \times \{0, 1\}^{l_d}\rightarrow \{0, 1\}^{l_r}$ be a PRF family, $ctr \in \{0,1\}^{l_r}$ be a counter.
 Each tag $T_i \in \Tt$, $1 \leq i \leq l$, is assigned with an unique identity $\mathit{ID}_i$,
a secret-key $k_i \in_R \{0, 1\}^{l_k}$, a counter $ctr_i$  with initial value 1.
For each tag $T_i \in \Tt$, the reader $R$ pre-computes an initial index $I_i =
 F_{k_i}(ctr_i||pad)$, and stores  $rcd_i = (I_i, k_i, ctr_i, ID_i)$ in its
database $\mathit{DB}^1$. The whole system's public parameters are  $para  =  \{\kappa, l, l_k, l_d, l_{p}, l_{v}, l_r, l_u, F, pad\}$.

\vspace{2mm}

\subsubsection{$\func{Protocol}\ \pi(R, T_i)$} The protocol runs as below.

%\begin{enumerate}[label=(\roman*)]
\begin{itemize}[]
\item $R$ chooses $r_R \in_R \{0, 1\}^{l_u}$,  and sends $c_1 = r_R$ to $T_i$.

\item  Upon receiving $c_1$ from $R$, $T_i$ chooses $r_T \in_R \{0, 1\}^{l_{v}}$,  computes $\alpha_{11} = F_{k_i}(ctr_i||pad)$ and  $\alpha_{13} = F_{k_i}(c_1||\alpha_{11}||\alpha_{12}) \oplus  ctr_i$, and  sets $\alpha_{12} = r_T$. Then $T_i$ sends $\alpha_1 = \{\alpha_{11}, \alpha_{12}, \alpha_{13}\}$ to $R$,  and updates its counter $ctr_i = ctr_i + 1$.
\item   After receiving $\alpha_1 = \{\alpha_{11}, \alpha_{12}, \alpha_{13}\}$, $R$ authenticates $T_i$ through the following steps.
 \begin{itemize}
   \item   Step 1: $R$ searches its database to find a record $rcd = (I, k,$ $ctr$, $\mathit{ID},$ $\mathit{PK}_T)$  such that $I = \alpha_{11}$ and $ctr = F_k(r_R||\alpha_{11}||\alpha_{12}) \oplus \alpha_{13}$. If it discovers one, it updates $ctr = ctr+1$ and $I= F_k(ctr||pad)$, and proceeds to Step 3; else, proceeds to Step 2.
       \vspace{1mm}
   \item  Step 2:  $R$ searches its database to find a record  $rcd = (I', k, ctr, ID, \mathit{PK}_T)$  such that $I = F_k(ctr'||pad)$ where $ctr' = F_k(r_R||\alpha_{11}||\alpha_{12}) \oplus \alpha_{13}$. If it discovers one, it updates $ctr  = ctr'+1$  and $I'= F_k(ctr||pad)$, and proceeds to Step 3; else, outputs $o_R = 0$.
         \vspace{1mm}
   \item   Step 3: $R$ computes $c_2 = F_{k}(r_R||ctr||\alpha_{12})$, sends $c_2$ to $T_i$, and outputs $o_R = 1$.
 \end{itemize}
\item Upon receiving $c_2$, if $c_2=F_{k_i}(c_1||ctr_i||r_T)$, $T_i$ outputs $o_{T_i}=1$; else,  outputs  $o_{T_i}=0$.
% \end{enumerate}
\end{itemize}

We summarize the security analysis results on the  protocol $\pi$ with Theorem \ref{theorem:munp} while omitting  the proofs, as they are similar to the ones provided in \cite{Li2011,Deng2011}.

\vspace{1mm}
\begin{theorem} \label{theorem:munp}
If the function family $F: \{0, 1\}^{l_k} \times \{0, 1\}^{l_d} \rightarrow \{0, 1\}^{l_r}$ is a PRF family, then
the RFID system $\mathit{MA} = (R, \Tt, \func{Setup}, \pi)$ defined above satisfies adaptive completeness, mutual authentication, and unp$^\#$-privacy.
\end{theorem}

\subsection{An RFID mutual authentication protocol with PoP}

We now construct a system   $\mathit{MAPoP}=(R, \Tt, \func{Setup^*}, \pi^*,$ $\func{CredGen}, \func{CredVeri})$ with PoP based on the system   $\mathit{MA} = (R, \Tt, \func{Setup}, \pi)$. %as it is lightweight and suitable for low-cost devices.
Below we show the additions to  $RS$.

\smallskip
%\subsubsection{$\func{Setup^*}(\kappa,l)$}  We add a cryptographic hash function   $H: \{0, 1\}^* \rightarrow\{0, 1\}^{l_r}$,  another PRF family $G: \{0, 1\}^{l_k} \times \{0, 1\}^{2l_r}\rightarrow \{0, 1\}^{2l_r}$, and the elliptic curve-based Schnorr signature scheme \cite{Schnorr91}.
%The Schnorr signature scheme is initialized with parameters $(E(\Ff_q),P, t,h)$, where $P$ is a point on the elliptic curve $E(\Ff_q)$ with order $t$,  $\Ff_q$ is a finite field with order $q$, and  $h()$ is a hash function such that $h: \{0,1\}^* \rightarrow \Zz_q$. %The key generation, signing, and verification algorithms are denoted as $\mathcal{K}^{Schn}$, $\mathcal{S}^{Schn}$,  and $\Vv^{Schn}_{\mathit{PK}}$ respectively.
% Each tag $T_i$ and the reader are assigned with a public and secret key pair for the signature scheme, denoted as $(\mathit{PK}_{T_i}, sk_{T_i})$ and  $(\mathit{PK}_{R}, sk_{R})$ respectively. Each tag $T_i$ is also assigned with another secret key  $k'_{T_i} \in_R \{0,1\}^{l_k}$. In the database $\mathit{DB}^*$, each record $rcd_i^*$ stores $(I_i, k_i, k'_i, ctr_i, ID_i, \mathit{PK}_{T_i})$.
%The whole system's public parameters  $para^*$ additionally include   $\{H, \mathit{PK}_{T_1},  \cdots, \mathit{PK}_{T_l}, \mathit{PK}_R, E(\Ff_q), P, t,h, G\}$.

\subsubsection{$\func{Setup^*}(\kappa,l)$}  We add a cryptographic hash function   $H: \{0, 1\}^* \rightarrow\{0, 1\}^{l_r}$,  another PRF family $G: \{0, 1\}^{l_k}$ $\times \{0, 1\}^{2l_r}\rightarrow \{0, 1\}^{2l_r}$, and a  signature scheme $\Dd\Ss$.  $\func{Setup^*}$ $(\kappa,l)$ is the same as defined in Section \ref{sec:genConstr}.

\vspace{1mm}

\subsubsection{$\func{Protocol}$ $\pi^*(R, T_i)$}  
\begin{figure*}[htbp]
  \centering
  \setlength{\fboxrule}{0.5pt}
  \setlength{\fboxsep}{0cm}
  \fbox{\includegraphics[scale=0.56]{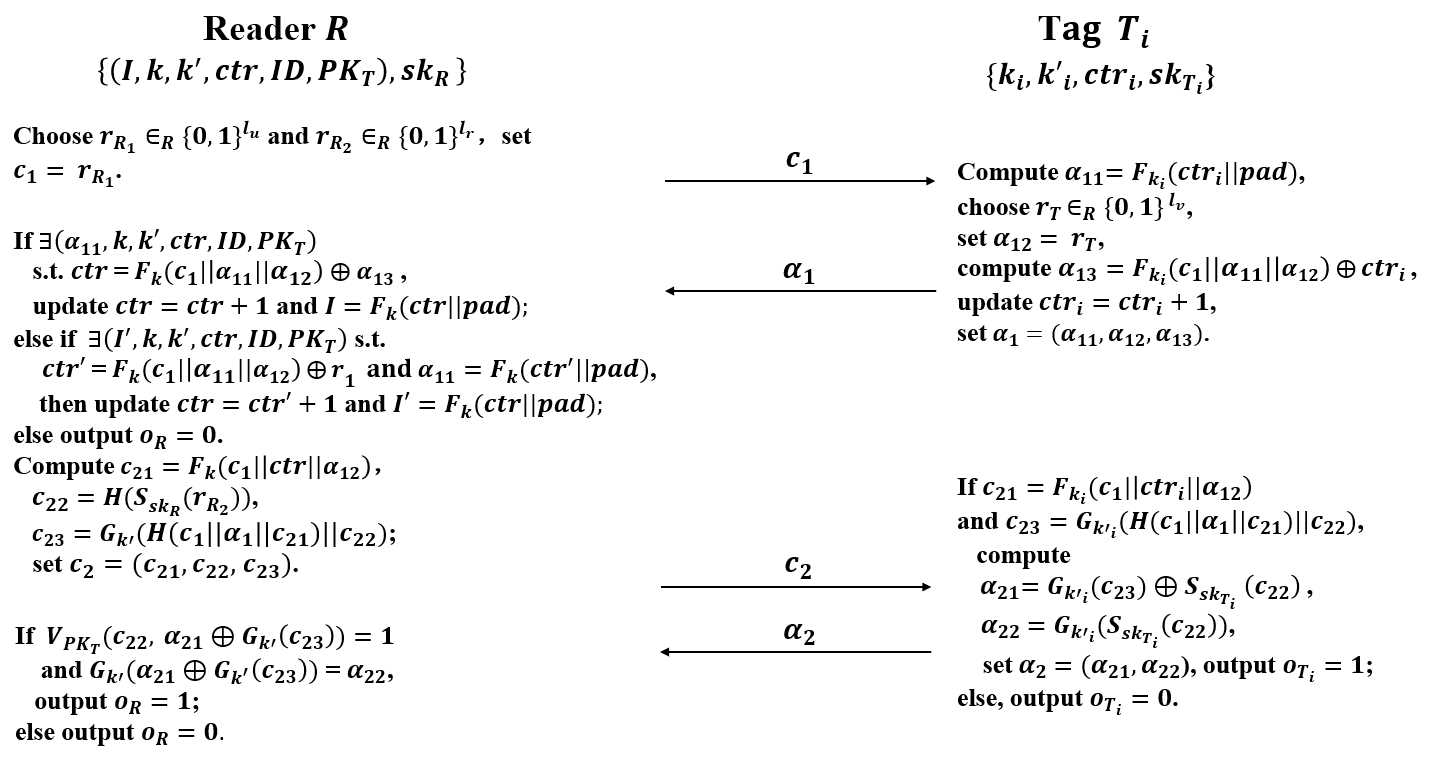}}\\
  \caption{An RFID Mutual Authentication Protocol With PoP}\label{fig:ourprotocol}
\end{figure*}

The  protocol is  depicted in Figure \ref{fig:ourprotocol}, and runs as follows.

\begin{itemize}
  \item  $R$ chooses ${r_R}_1\in_R \{0, 1\}^{l_u}$,  ${r_R}_2 \in_R$ $\{0,1\}^{l_r}$   and sends $c_1 = {r_R}_1$ to $T_i$.
  \item  Upon receiving $c_1$, the tag $T_i$ chooses ${r_T}\in_R$ $\{0,1\}^{l_{v}}$, computes $\alpha_{11} = $ $F_{k_i}(ctr_i||pad)$,   $\alpha_{12} = {r_T}$,  and $\alpha_{13}= F_{k_i}(c_1||\alpha_{11}||\alpha_{12}) \oplus  ctr_i$. Then $T_i$ sends $\alpha_1 = \{\alpha_{11}, \alpha_{12}, \alpha_{13}\}$ to the reader $R$, and updates its counter $ctr_i = ctr_i + 1$.
  \item  After receiving $\alpha_1 = \{\alpha_{11}, \alpha_{12}, \alpha_{13}\}$, the reader $R$ authenticates the tag through the following steps.
 \begin{itemize}
   \item  Step 1: $R$ searches its database to find a record  $rcd = (I, k, k',$ $ctr$, $\mathit{ID},$ $\mathit{PK}_T)$ such that $I = \alpha_{11}$ and $ctr = F_k({r_R}_1||\alpha_{11}||\alpha_{12}) \oplus \alpha_{13}$. If  it discovers   one,  it updates $ctr = ctr+1$  and $I= F_k(ctr||pad)$, and  proceeds to Step 3; else, proceeds to Step 2.
 %      \vspace{2mm}
   \item   Step 2: $R$ searches its database to find a record
        $rcd = (I', k, k', ctr, ID, \mathit{PK}_T)$  such that $I' = F_k(ctr'||pad)$ where $ctr'$ $=$ $F_k(c_1||$ $\alpha_{11}||$$\alpha_{12})$ $\oplus$ $\alpha_{12}$. If it discovers   one,  it updates $ctr  = ctr'+1$ and $I'= F_k(ctr||pad)$, and  proceeds to Step 3; else, outputs $o_R = 0$.
   \item  Step 3:  $R$ computes $c_{21}$ $=$ $F_{k}(r_{R1}||ctr||\alpha_{12})$, $c_{22}$ $=$ $H(\Ss_{sk_R}(r_{R2}))$, and  $c_{23}$ $=$ $G_{k'}(H(c_1||\alpha_1||c_{21})||c_{22})$. Then $R$ sends  $c_2$ $=$ $(c_{21},c_{22},c_{23})$  to $T_i$.

 \end{itemize}
  \item  Upon receiving $c_2$, if  $c_{21}=F_{k_i}(c_1||ctr_i||\alpha_{12})$ and  $c_{23} = G_{k'_i}(H(c_1||\alpha_1||c_{21})||c_{22})$, $T_i$ computes  $\alpha_{21} = G_{k_{T_i}'}(c_{23})$ $\oplus \Ss_{sk_{T_i}}(c_{22})$, $\alpha_{22} = G_{k_{T_i}'}(\Ss_{sk_{T_i}}(c_{22}))$ sends $\alpha_2 =(\alpha_{21}, \alpha_{22})$ to $R$, and outputs $o_{T_i}=1$; otherwise it  outputs $o_{T_i}=0$.
  \item  Upon receiving $\alpha_2$, if $\Vv_{\mathit{PK}_{T_i}}$ $(c_{22},$ $\alpha_{21}$ $\oplus$ $G_{k'}(c_{23})) = 1$ and $G_{k'}(\alpha_{21} \oplus G_{k'}(c_{23})) = \alpha_{22}$, $R$ outputs $o_R = 1$; else,  outputs $o_R = 0$.
\end{itemize}

\smallskip

 $\func{CredGen}(para, sk_R, s_R^j, {DB^j})$: Suppose the reader $R$ has finished its $j$-th session of the protocol $\pi^*$, and  identified the tag as $T_i$.
   $\func{CredGen}$ computes a PoP credential $cred$ $=$ $\{R,$ $T_i,$ $r_{R_2},$ $\Ss_{sk_R}(r_{R_2})$, $\alpha_{21} \oplus G_{k'_i}(c_{23})\}$, where $r_{R_2}$, $\alpha_{12}$, and $\alpha_2$ are stored in $s_R^j$, and  $k'_i$ comes from $T_i$'s record $rcd_i \in DB^j$.

\smallskip

  $\func{CredVeri}(para, cred)$: Given a PoP credential $cred$ $=\{R,T_i, cred_1, cred_2, cred_3\}$ and the system's public parameters $para$, it outputs the verification result ``1'' if $\Vv_{\mathit{PK}_R}(cred_1, cred_2) = 1 \wedge \Vv_{\mathit{PK}_{T_i}}(H(cred_2), cred_3) = 1$, and outputs ``0'' in any other cases.

\vspace{2mm}

We summarize the security analysis results on the RFID system $\mathit{MAPoP}$ with Theorem \ref{theo:mapop} (proof sketches  are provided in Appendix).

\smallskip
\begin{theorem} \label{theo:mapop}
If both the function family  $F: \{0, 1\}^{l_k} \times \{0, 1\}^{l_d} \rightarrow \{0, 1\}^{l_r}$ and the function family $G:\{0,1\}^{l_k}\times\{0,1\}^{2l_g}\rightarrow \{0,1\}^{2l_g}$ are PRF families, and the  signature scheme $\Dd\Ss$ is complete and satisfies EU-CMA, then $\mathit{MAPoP}$ satisfies adaptive completeness, mutual authentication, PoP credential unforgeability, and unp$^\#$-privacy under the random oracle model.
\end{theorem}

\vspace{-2mm}
\subsection{Performance} \label{sec:performance}

We consider the implementation for $\mathit{MAPoP}$ with 128-bit security.
Schnorr$\mathbb{Q}$ \cite{Costello16} is a digital signature scheme that is based on the well-known Schnorr signature
scheme \cite{Schnorr91} combined with the use of the elliptic curve Four$\mathbb{Q}$ \cite{Costello15}. Schnorr$\mathbb{Q}$ offers extremely fast, high-security digital signatures targeting the 128-bit security level. We also consider SEMECS \cite{Yavuz19}, a $K$-time ECC-based signature scheme that is more  efficient compared to its counterparts. It achieves optimal signature and private key sizes, and requires constant-size storage and no ECC operation at the signer.  However, there are some tradeoffs.  SEMECS only supports generating $K$ signatures at most for each secret key, and the public key's size is linear to $K$.
Although in our generic  construction, the reader and the tags adopt the same signature scheme, it is also feasible that the reader adopts  a full-time signature  and the tags adopt a $K$-time  one, as long as   the full-time signature scheme satisfies  completeness and EU-CMA,  and the $K$-time signature scheme satisfies  completeness and $K$-time EU-CMA defined in \cite{Yavuz19}.

We denote SEMECS using  Four$\mathbb{Q}$   as SEMECS$\mathbb{Q}$. We select  Schnorr$\mathbb{Q}$ and SEMECS$\mathbb{Q}$ as candidate signature schemes for $\mathit{MAPoP}$.
We select BLAKE3 \cite{BLAKE3} to instantiate the PRF families $F$ and $G$  and serve as the underlying hash function for $\mathit{MAPoP}$  due to its high efficiency and high security.  BLAKE3  is a 128-bit secure cryptographic hash which can operate in three modes with a single algorithm. We  use  $\mathrm{hash}$ mode and $\mathrm{keyed\_hash}$ mode in our implementation.

We consider three different kinds of instantiations for $\mathit{MAPoP}$, denoted as IMP1, IMP2, and IMP3 respectively. They are all based on BLAKE3,  but differ in the selection/implementation of the signature scheme.
In IMP1, both the reader and the tags adopt Schnorr$\mathbb{Q}$; and the tags need to conduct ECC point multiplication operations when generating signatures.  In IMP2, both the reader and the tags adopt Schnorr$\mathbb{Q}$; and the tags use pre-computed $(r, rP)$
 pairs\footnote{$P$ is a point on Four$\mathbb{Q}$ with order $t$, and  $r \in_R [1,t]$. } for generating signatures. In IMP3, the reader adopts Schnorr$\mathbb{Q}$, while the tags adopt  SEMECS$\mathbb{Q}$ which does not require any ECC operation for signing a message. We provide performance analysis of IMP1, IMP2, IMP3, and  $MA$ (without PoP).  The performances of  $MA$ based on BLAKE3 serve as a baseline for the performances of IMP1, IMP2, and IMP3.

\subsubsection{Storage requirement}
For the reader, we only list the size of each tag's record in the database. In $\mathit{MAPoP}$,  the reader stores  $rcd_i = \{I_i, k_{T_i}, k'_{T_i}, ctr_i, ID_i, PK_{T_i}\}$ for each tag $T_i$; and $T_i$ stores $\{k_{T_i}, k'_{T_i}, ctr_i, sk_{T_i}\}$. In  $\mathit{MA}$, each reader stores  $rcd_i = \{I_i, k_{T_i}, ctr_i, ID_i\}$ for each tag $T_i$; and  $T_i$ stores $\{k_{T_i}, ctr_i\}$.

 Schnorr$\mathbb{Q}$'s secret key size is 32 bytes,  public key size is 32 bytes, and signature size is 64 bytes. The $K$-time signature scheme SEMECS$\mathbb{Q}$'s secret key size is 32 bytes, signature size is  32 bytes,  and public key size is $64 + 64K + |K|$ bytes. % $64+64\times K+|K|$
BLAKE3 takes an input of any byte length in $[0,2^{64}]$ (and together with a 32-byte key  in $\mathrm{keyed\_hash}$ mode  only),  and by default generates a 32-byte output. The output can be extended to any byte length in $[0,2^{64}]$. % in both  of $\mathrm{hash}$ mode and $\mathrm{keyed\_hash}$ mode
 According to BLAKE3's parameter settings, $k_{T_i}$, $k'_{T_i}$, $I_i$, and $ctr_i$ are all 32 bytes. We set $ID_i$ as 32 bytes,  as it is the maximum length of an electronic product code (EPC) \cite{EPCglobal}. In IMP2, we assume that a tag stores $K$ pairs of pre-computed values $(r, rP)$.

The storage requirements  are
shown in Table \ref{table:comp}. For IMP2 and IMP3, we set $K = 2^{17}$, which  allows a tag to run a full protocol session in every 20 minutes for 5 years without updating pre-computed values/secret key. Note that, both the reader and the tags  need to store the value of $K$  and maintain a counter for the signatures which require $2|K|$ bytes, while we omit to count them, as $|K|$ only takes three bytes.

\begin{table}[!ht]
  \centering
  \caption{Comparison on the Storage Requirements \\and Communication Overheads} \label{table:comp}
  %\begin{threeparttable}[b]
  \scalebox{0.9}{
  \begin{tabular}{|c|c|c|c|c|}
             \hline
             % after \\: \hline or \cline{col1-col2} \cline{col3-col4} ...
                & IMP1 & IMP2 & IMP3& $\mathit{MA}$\\
              \hline
               \multicolumn{5}{|l|}{storage requirement in terms of bytes} \\
               \hline
              $R$ & 192  & 192  & $\approx$ 8 MB &96 \\
              \hline
              $T_i$ &  128 & $\approx$ 8 MB  & 128   & 64\\
             \hline
             \multicolumn{5}{|l|}{message size in terms of bytes} \\
               \hline
               1st round & 32 & 32 & 32  & 32\\
               \hline
               2nd round & 96 & 96 & 96 & 96\\
               \hline
               3rd round & 96 & 96 & 96 & 32 \\
               \hline
               4th round & 96 & 96 & 64 & -\\
             \hline
           \end{tabular}}
%  \begin{tablenotes}
%  \end{tablenotes}
%  \end{threeparttable}
\end{table}

\subsubsection{Communication requirement}
The  communication overheads are also shown in Table \ref{table:comp}.  IMP3's fourth round message is 32 bytes shorter than IMP1's and  IMP2's due to the following reason. The forth round message of $\pi^*$ consists of two parts,  $\alpha_{21}$ and $\alpha_{22}$, where  $\alpha_{21} = G_{K'_i}(c_{23}) \oplus \Ss_{sk_{T_i}}(c_{22})$.  In  IMP1 and IMP2, because  Schnorr$\mathbb{Q}$'s  signature size is 64 bytes, then   $G_{K'_i}$ is instantiated using  BLAKE3 with an extended 64-byte output   and   $\alpha_{21}$ is 64 bytes. While in IMP3, because SEMECS$\mathbb{Q}$'s  signature size is 32 bytes,  then $G_{K'_i}$ is instantiated using  BLAKE3 with a default 32-byte output  and  $\alpha_{21}$ is 32 bytes.

\subsubsection{Computational requirement}
Let $\mathit{eMul}$, $eAdd$, $\mathit{mMul}$, and $H$ denote  an ECC point multiplication operation, an ECC point addition operation, a modular multiplication operation,
 and a hash function operation respectively; and $T_{\mathit{eMul}}$, $T_{eAdd}$, $T_{\mathit{mMul}}$, and $T_{H}$ denote their running time  respectively.
Generally, an ECC point multiplication operation requires much more time than other operations.
 For example, under the settings of \cite{Chatterjee14},  $T_{\mathit{eMul}} \approx 3,333 T_{H}$,  $T_{\mathit{eAdd}}  \approx
14T_{H}$, and $T_{mMul} \approx 2.7T_H$.

$\mathsf{Sync}$ denotes that the reader is synchronized with a tag, and  can identify the tag using the index $I$. $\mathsf{Desync}$ denotes that  the reader is de-synchronized  with a tag, so that it  needs  to search the whole database with $l$ records for identifying the tag, where $l$ is the number of  tags. We omit  cheap operations if there exists at least one expensive one. In  Schnorr$\mathbb{Q}$, generating a signature incurs one $H$ and one $\mathit{mMul}$, and verifying a signature incurs two $\mathit{eMul}$. In  SEMECS$\mathbb{Q}$, generating a signature incurs three $H$ and one $\mathit{mMul}$, and verifying a signature incurs two $\mathit{eMul}$.
The  computational overheads of IMP1, IMP2, IMP3, and $\mathit{MA}$ are
shown in Table \ref{table:compcmp}.

\begin{table}[!htbp]
\centering
\caption{Comparison on Computational Overheads} \label{table:compcmp}
%  \begin{threeparttable}[b]
  \scalebox{0.9}{
\begin{tabular}{|c|c|c|c|c|}
\hline
  &IMP1 & IMP2 & IMP3 & $\mathit{MA}$\\
 \hline
$R$ ($\mathsf{Sync}$) &   3$\mathit{eMul}$ &  3$\mathit{eMul}$ &  3$\mathit{eMul}$ & 3$H$   \\
\hline
$R$ ($\mathsf{Desync}$) &  $O(l) \cdot H$  &  $O(l) \cdot H$ & $O(l) \cdot H$ &$O(l) \cdot H$  \\
  &+ 3$\mathit{eMul}$ &  + 3$\mathit{eMul}$ & +  3$\mathit{eMul}$ &\\
\hline
$T_i$  &  $\mathit{eMul}$ &        8$H$&    10$H$  &  3$H$    \\
   & & + $\mathit{mMul}$&   + $\mathit{mMul}$ &  \\
\hline
\end{tabular}}
% \begin{tablenotes}
%
%  \end{tablenotes}
% \end{threeparttable}
\end{table}

\subsubsection{Discussions} None  of  IMP1, IMP2, and IMP3 outperforms the other two in all aspects.
IMP1 has the highest requirements on tag computational capability. Although Tan et al.  proposed a 163-bit ECC processor \cite{Tan17} suitable for UHF RFID tags, it only supports 80-bit security. There are efficient hardware implementations for 256-bit ECC (e.g., \cite{Rahman19,Kudithi20}), however,  it may still require further development to make them applicable for RFID tags. IMP1 also has the highest computational overheads on the tag side due to the expensive $\mathit{eMul}$.  % The tags need to conduct expensive $eMul$ operations.
Tan et al. showed that their ECC processor supports the running of their authentication protocol up to 1,500 sessions per minute. As both their authentication protocol and IMP1 require one $\mathit{eMul}$ on the tag side in each session, we estimate that our generic construction (with 80-bit security) can be executed with comparable overhead if such ECC processor is implemented on the tag side.

Among IMP1, IMP2, and IMP3, IMP2 has the highest requirement on tag storage.  Each pair of   $(r, rP)$ takes 64 bytes.  Besides leaving 96 bytes for storing keys,  a tag with 2048-byte user memory (e.g., \cite{nxp}) can only store 30 pairs of  $(r, rP)$. As each session of   $\pi^*$ consumes one pair of $(r, rP)$, 30 pairs may not be always sufficient in practice.

Among IMP1, IMP2, and IMP3, IMP3 has the highest requirement  on reader storage. If $K  = 2^{17}$, it takes 8 TB for the reader to store one million tags' public keys. It is not difficult to equip the reader's back-end server to match this storage requirement. %We estimate that running a  session of $\pi^*$ in IMP3 would take several times of the time for running a session of $\pi$ in $\mathit{Ma}$.
Since IMP3 does not involve any expensive operation on the tag side, we consider IMP3 as the most applicable instantiation of $\mathit{MAPoP}$ among the three options.

%!TEX root = main.tex
\vspace{-1mm}
\section{Summary and Future Works}
Proof of possession (PoP) is a property critical for many RFID-enabled applications, especially for RFID data-driven applications. We are the first to propose a formal framework for RFID tag/mutual authentication with PoP. We provided a secure and privacy-preserving generic construction  in our framework based on any secure and unp$^\#$-privacy-preserving tag/mutual authentication protocol, an EU-CMA secure signature scheme, a cryptographic hash function, and a PRF. We also proposed a concrete construction for RFID mutual authentication protocol with PoP,  conducted security analysis, and provided a practical implementation solution.  Potential future research directions include exploring other (stronger or weaker) formal frameworks for RFID tag/mutual authentication with PoP  and designing RFID tag/mutual authentication protocols that support PoP  with better practicality.

\bibliographystyle{plain}

\end{document}

\title{Appendix}

%\author{\IEEEauthorblockN{Shaoying Cai, %\IEEEauthorrefmark{1},
%Yingjiu Li, %\IEEEauthorrefmark{2},
%Changshe Ma, %\IEEEauthorrefmark{3},
%Sherman S. M. Chow, %\IEEEauthorrefmark{4}, and
%Robert H. Deng~\IEEEmembership{Fellow,~IEEE}} %\IEEEauthorrefmark{5},
%%
%%\IEEEauthorblockA{\IEEEauthorrefmark{1}College of Computer Science and Electronic Engineering, Hunan University, China}
%%\IEEEauthorblockA{\IEEEauthorrefmark{2}Department of Information Engineering, The Chinese University of Hong Kong, Hong Kong}
%%\IEEEauthorblockA{\IEEEauthorrefmark{3}School of Computer, South
%%China Normal University, China}
%% \IEEEauthorblockA{\IEEEauthorrefmark{4}Computer and Information Science Department, University of Oregon, USA}
%% \IEEEauthorblockA{\IEEEauthorrefmark{5}School of Information Systems, Singapore Management University, Singapore}
%%\thanks{%Manuscript received December 1, 2012; revised August 26, 2015.
%%S. Cai is with the College of Computer Science and Electronic Engineering, Hunan University, China.
%%Y. Li is with the Computer and Information Science Department, University of Oregon, USA.
%%C. Ma is with the School of Computer, South China Normal University, China.
%%S. Chow is with the Department of Information Engineering, The Chinese University of Hong Kong, Hong Kong.
%%R. H. Deng is with the School of Computing and Information Systems, Singapore Management University, Singapore.
%%}
%}
%
%\markboth{}%Journal of \LaTeX\ Class Files,~Vol.~14, No.~8, August~2015%
%{Shell \MakeLowercase{\textit{et al.}}: Bare Demo of IEEEtran.cls for IEEE Communications Society Journals}

\maketitle
%
%\input{0-abstract}
%
%\input{1-intro}
%
%\input{2-related}
%
%\input{3-unptau}
%\input{4-systemModel}
%
%
%\input{5-genericSolution}
%\input{6-Protocol}
%
%\input{7-conclusion}
%
%\input{8-ref}

%!TEX root = main.tex

%\appendices
%\section{Proofs}
%\label{sec:proofs}

\section{Proofs for Theorem 1}

Note that $\pi'(R,T_i)$ is a mutual authentication protocol but not an identification protocol. We assume that before $R$ and $T_i$ run a session of $\pi'$,   $R$ has known the identity of $T_i$. For example, in our generic construction, $R$ can identify $T_i$  through  running a session of $\pi$.

\smallskip
Obviously, Theorem 1 holds if all of Lemma \ref{lem:pipc}-\ref{lem:pipunp} hold.

\smallskip
\begin{lem} \label{lem:pipc}
If  the function family $G:\{0,1\}^{l_k} \times \{0,1\}^{2l_g} \rightarrow \{0,1\}^{2l_g}$ is a PRF family, and the underlying signature is complete, then $\pi'(R, T_i)$  is complete.
\end{lem}
\begin{IEEEproof} Obviously, this lemma holds.  \end{IEEEproof}

\smallskip
\begin{lem} \label{lem:pipmutual}
If the function family $G:\{0,1\}^{l_k} \times \{0,1\}^{2l_g} \rightarrow \{0,1\}^{2l_g}$ is a PRF family, then $\pi'(R, T_i)$ satisfies mutual authentication.
\end{lem}

\begin{IEEEproof}(Sketch) To prove that $\pi'(R, T_i)$ achieves mutual authentication, we first construct a protocol $\pi''(R, T_i)$. The protocol $\pi''(R, T_i)$ is the same as $\pi'(R, T_i)$ except that $c' = r \in_R \{0, 1\}^{l_g}$ and $\alpha' = G_{k'_{T_i}}(c'') \oplus y$, where $y \in_R \{0,1\}^{2l_g}$; while in $\pi'(R, T_i)$, $c' = H(\Ss_{PK_R}(r))$ and $\alpha' = G_{k'_{T_i}}(c'')\oplus y$,  where $y = \Ss_{sk_{T_i}}(c')$.

We prove that if the function family $G:\{0,1\}^{l_k} \times \{0,1\}^{2l_g} \rightarrow \{0,1\}^{2l_g}$ is a PRF family, then $\pi''(R, T_i)$ achieves mutual authentication. If there is an adversary $\Aa$ $(\epsilon,t, n_1,n_2,n_3,n_4)$-breaks  the mutual authentication of  $\pi''(R, T_i)$, then we can construct algorithm $\Bb$ that can pass the $PTPT$ for the function family $G$ using $\Aa$ as a subroutine.

Algorithm $\Bb$ constructs a system $RS=(R, \Tt, \func{Setup}, \pi'')$ which contains $l$ tags. Then it chooses $i \in_R [1, l]$.  Let $\Tt' = \{\Tt-T_i\}$.  For each tag $T_j \in \Tt'$, %($1 \leq j \leq l$ and $j \neq i$)
algorithm $\Bb$ chooses a secret key $k'_{T_j} \in_R \{0,1\}^{l_k}$. And $T_i$ has a key $k$ which is unknown to  algorithm $\Bb$, where $k \in_R \{0,1\}^{l_k}$.
  Algorithm $\Bb$  answers $\Aa$'s queries related to $T_j \in \Tt'$ using the key $k'_{T_j}$, and answers the queries related  to $T_i$ with an oracle $O_G$. Upon queried with a $2l_g$-bit string $x$, the oracle $O_G$ returns $y = G_k(x)$ if $b=1$,  or $y \in_R \{0,1\}^{2l_g}$ if $b=0$.

In the learning stage, when adversary $\Aa$ queries $O_2$, $O_3$ and $O_4$ about the tags in $\Tt'$, algorithm $\Bb$ uses
the tags' keys to generate responses according to the protocol $\pi''$'s procedures. When $\Aa$ queries $O_4$ about
$T_i$, algorithm $\Bb$ aborts. If $\Aa$ queries $O_2$ or $O_3$ about $T_i$, $\Bb$ uses $O_G$ to generate replies. %When $\Aa$ submits transcripts of a session $trs$, if $trs$ are not related to $T_c$, $\Bb$ aborts.

Finally, $\Aa$ stops and submits transcripts $trs = \{(c', c''),$ $(\alpha', \alpha'')\}$ of a session which are different from any existing session's transcripts.  If $trs$ are not related to $T_i$, $\Bb$ aborts, and randomly outputs a bit.  $trs$ are related to $T_i$ are considered valid if $c'' = O_G(trs||c')$ and $\alpha_{\gamma+1}'' = O_G(O_G(c'')\oplus\alpha')$.
Then $B$ outputs $b'=1$  if $trs$ are valid; else outputs 0.

Note that if $O_G$ answers queries with random values (when $b=0$), in any case the probability that $\Aa$ can provide valid transcripts which are related to $T_i$ is negligible. If $O_G$ answers queries using $G_k$ (when $b = 1$), then if $\Aa$ $(\epsilon,t, n_1,n_2,n_3,n_4)$-breaks the mutual authentication of $\pi''$,  the probability that $\Aa$ can provide valid transcripts which are related to $T_i$ is $\frac{\epsilon}{l}$.
Thus if $\Aa$ $(\epsilon,t, n_1,n_2,n_3,n_4)$-breaks the mutual authentication of $\pi''$, where $\epsilon$ is non-negligible, then $\Bb$ can pass the \emph{PTPT} with non-negligible advantage $\frac{\epsilon}{l}$, and the running time of $\Bb$ is approximate to that of $\Aa$.

%Now we show that if $\pi''(R, T_i)$ achieves mutual authentication, then $\pi'(R, T_i)$ achieves mutual authentication.
The mutual authentication procedures of $\pi''(R, T_i)$ are the same as $\pi'(R, T_i)$ essentially. For reader authentication, given any $trs \in \{0,1\}^{l_\pi}$, for any $c' \in \{0,1\}^{l_g}$, the reader can compute a valid $c'' =G_{k'_{T_i}}(trs||c')$, while any one without $k'_{T_i}$ cannot provide any valid pair of $(c', c'')$. Thus in $\pi'(R, T_i)$, changing the way of generating $c'$ does not effect the reader authentication  essentially. Similarly, changing the way of generating $y$ in $\pi'(R, T_i)$ does not effect the tag authentication  essentially either. Thus  if $\pi''(R, T_i)$ archives mutual authentication, then $\pi'(R, T_i)$ achieves mutual authentication.\end{IEEEproof}

\smallskip
\begin{lem} \label{lem:pipunp}
If the function family $G:\{0,1\}^{l_k} \times \{0,1\}^{2l_g} \rightarrow \{0,1\}^{2l_g}$ is a PRF family, then $\pi'(R, T_i)$ achieves unp$^\#$-privacy under the random oracle model.
\end{lem}

\begin{IEEEproof}(Sketch) If  an adversary $\Aa$
can $(\epsilon, t, n_1, n_2,n_3,$ $n_4, n_5)$-break the unp$^\#$-privacy of $\pi'(R, T_i)$, we can construct an algorithm $\Bb$
that can pass the $PTPT$ for the function family $G$ using $\Aa$ as a subroutine.

Algorithm $\Bb$ sets up an experiment Exp$_{\Aa}^{Unp^\#}$ for $\Aa$. Then $\Bb$ prepares a reader $R$ and a set of $l$ tags $\Tt$. Each tag $T_j$ $(1\leq j\leq l)$ and $R$ are assigned with a pair of public and secret keys, denoted as $(PK_{T_j}, sk_{T_j})$ and $(PK_{R}, sk_{R})$ respectively.  Now $\Bb$ selects an index $i$ between $1$ and $l$ randomly. The key
of $T_i$ is implicitly set to be
$k$, which is unknown to $\Bb$. However, $\Bb$ can access an oracle $O_G$  for a function $f = G_k$   or $f \in_R \mathsf{Rand}^{\Dd\rightarrow\Rr}$. For $1 \leq j \leq l$ and $j \neq i$, $\Bb$ selects a random key
$k'_{T_j} \in_R \{0,1\}^{l_k}$, then sets the key of the tag $T_j$ as $k'_{T_j}$.

\emph{In the learning stage.} When adversary $\Aa$ queries $O_2$, $O_3$ and $O_4$ about the tags in $\Tt'$, algorithm $\Bb$ uses
the tags' keys to respond according to the protocol $\pi'$'s procedures. When $\Aa$ queries $O_2$ or $O_3$ about $T_i$, $\Bb$ uses $O_f$  and $sk_{T_i}$ to generate replies. When $\Aa$ queries $O_4$ about
$T_i$, $\Bb$ aborts and randomly outputs a bit. Besides, $\Bb$ needs to maintain query histories of $\Aa$ and provide  protocol execution results as we defined in Exp$_{\Aa}^{Unp^\#}$ when $b =0$. Finally, when $\Aa$ submits an uncorrupted challenge tag $T_c$, if $T_c \neq T_i$, $\Bb$ aborts and randomly outputs a bit; else, proceeds to the guess stage.

\emph{In the guess stage.} Algorithm $\Bb$ answers adversary $\Aa$'s queries on $O_2$ and
$O_3$ about $T_c$ using $O_G$. Similarly as in the learning stage, $\Bb$ needs to maintain query  histories of $\Aa$ and provide protocol execution results as we defined in Exp$_{\Aa}^{Unp^\#}$ when $b =0$.

\emph{Output. }Finally, adversary $\Aa$ outputs a bit $b'$. $\Bb$ also takes $b'$ as its output.
If $\Bb$ does not abort during the simulation, $\Bb$'s simulation is perfect and is identically
distributed as the real one from the construction. It is obvious that the probability that
$\Bb$ does not abort during the simulation is $\frac{1}{l}$. Therefore, the advantage of $\Bb$ is at least $\frac{\epsilon}{l}$.
The running time of $\Bb$ is approximate to that of $\Aa$. This completes the proof. \end{IEEEproof}

\section{Proofs for Theorem 2}

 Theorem 2 holds if all of Lemma \ref{lem:pistarc}, Lemma \ref{lem:pistarmutual}, Lemma \ref{lem:pistarp}, and Lemma \ref{lem:pistarcu} hold. Below we provide proofs for them.

\smallskip
\begin{lem} \label{lem:pistarc}
If $\pi(R, T_i)$ satisfies adaptive completeness and $\pi'(R, T_i)$ is complete, then $\pi^*(R, T_i)$ satisfies adaptive completeness.
\end{lem}

\begin{IEEEproof}(Sketch)
Given a session of $\pi^*$ between the reader $R$ and a valid tag $T_i$, let $E_{\pi^*_0}$, $E_{\pi^*_1}$, and $E_{\pi^*_2}$ denote the event that $o^{*}_R = 0$ , $R$ identifies  $T_i$ as a different tag $T_{j}$, and $o^{*}_{T_i} = 0$ respectively. Intuitively, to satisfy adaptive completeness, $Pr[E_{\pi^*_0}\vee E_{\pi^*_1} \vee E_{\pi^*_2}]$ should be negligible, which requires all of  $Pr[E_{\pi^*_0}]$, $Pr[E_{\pi^*_1}]$, and $Pr[E_{\pi^*_2}]$ to  be negligible.

The reader first authenticates $T_i$ according to $\pi$'s procedures and $o_R$ is the intermediate result. Note that the protocol $\pi$ could be a tag authentication protocol  or a mutual authentication protocol. We consider the case that $\pi$ is a tag authentication protocol. Let $E_{\pi_0}$, $E_{\pi_1}$, and $E_{\pi_3}$ denote the event that  $o_R = 0$, $R$ identifies  $T_i$ as a different tag $T_{j}$, and $R$ correctly identifies $T_i$  according to $\pi$'s procedures respectively. If  $o_R = 1$, then $R$ continues  to run the protocol $\pi'$'s procedures. Suppose that $R$ has authenticated $T_i$ as $T_j$ (if the authentication is correct, $j = i$; else, $j \neq i$),  $E_{\pi'_0}$ and $E_{\pi'_2}$ denote the event that $o_R' = 0$ and $o_{T_i}' = 0$  respectively when $R$ runs $\pi'$ with $T_i$ using $T_j$'s keys. The session of $\pi^*$ has final outputs $o_{T_i}^* = o_{T_i}'$ and $o_R^* = o_R'$.

We first compute the probability that the event $o^{*}_R = 0$ happens.
\begin{eqnarray} \label{eq1}
% \nonumber to remove numbering (before each equation)
 \nonumber Pr[E_{\pi^*_0}] &=& Pr[E_{\pi_0}] + Pr[E_{\pi'_0}|E_{\pi_1}] Pr[E_{\pi_1}] \\
 \nonumber  & & + Pr[E_{\pi'_0}|E_{\pi_3}] Pr[E_{\pi_3}] \\
 &\leq& Pr[E_{\pi_0}]+ Pr[E_{\pi_1}].
\end{eqnarray}

The probability that $R$ authenticates $T_i$ as another $T_j$ with $\pi^*$ should be lower than that with $\pi$ only, as $\pi^*$ contains another round mutual authentication procedures $\pi'$.
\begin{equation} \label{eq2}
 Pr[E_{\pi^*_1}] \leq Pr[E_{\pi_1}]
\end{equation}

Now we compute the probability that the event $o^{*}_{T_i} = 0$ happens.
\begin{eqnarray} \label{eq3}
% \nonumber to remove numbering (before each equation)
 \nonumber Pr[E_{\pi^*_2}] &=& Pr[E_{\pi'_2}|E_{\pi_1}] Pr[E_{\pi_1}] + Pr[E_{\pi'_2}|E_{\pi_3}] Pr[E_{\pi_3}]\\
 \nonumber &=& Pr[E_{\pi'_2}|E_{\pi_1}] Pr[E_{\pi_1}] + 0Pr[E_{\pi_3}] \\
 &\leq& Pr[E_{\pi_1}]
\end{eqnarray}

 As $\pi(R, T_i)$ satisfies adaptive completeness, $Pr[E_{\pi_0}]$, $Pr[E_{\pi_1}]$ are negligible. According to Equation \ref{eq1}-\ref{eq3},
 $Pr[E_{\pi^*_0}]$, $Pr[E_{\pi^*_1}]$, and $Pr[E_{\pi^*_2}]$ are all negligible. The analysis for the case that $\pi(R, T_i)$ is a mutual authentication can be conducted  similarly.
 Hence, we have proved that $\pi^*(R, T_i)$ achieves adaptive completeness.
\end{IEEEproof}

%\vspace{-4mm}
\smallskip
\begin{lem} \label{lem:pistarmutual}
If $\pi(R, T_i)$ achieves tag/mutual authentication and $\pi'(R, T_i)$ achieves mutual authentication, then $\pi^*(R, T_i)$ achieves mutual authentication.
\end{lem}
\begin{IEEEproof}(Sketch) %$\pi(R, T_i)$ and $\pi'(R, T_i)$ are two independent authentication protocols.
In a session of $\pi^*$, the reader first runs $\pi(R, T_i)$'s procedures. If it  authenticates $T_i$ as a valid tag, then continues to run the $\pi'$'s procedures using the $\pi$ part's transcripts $trs$ as inputs. According to Lemma \ref{Lemma1}, each run of $\pi$ would generate unique and unpredictable transcripts.
By feeding $\pi$'s  transcripts to $\pi'$, each session of $\pi^*(R, T_i)$ binds a session of $\pi(R, T_i)$ and a session of $\pi'(R, T_i)$. Then $\pi^*$ achieves mutual authentication if any of $\pi$ and $\pi'$ achieves mutual authentication.
 \end{IEEEproof}

\smallskip
\begin{lem} \label{Lemma1}
If $\pi(R, T_i)$ achieves adaptive completeness and tag/mutual authentication, the probability that the transcripts of any two sessions are the same is negligible.
\end{lem}
\begin{IEEEproof}(Sketch) If the transcripts of two sessions from two tags are the same, the reader may mis-authenticate each tag as the other. If the transcripts of two different sessions of a tag are the same, then the tag authentication/mutual authentication may fail.\end{IEEEproof}

\smallskip
\begin{lem} \label{lem:pistarp}
If $\pi(R, T_i)$ satisfies  unp$^\#$-privacy, and $\pi'(R, T_i)$ satisfies mutual authentication and unp$^\#$-privacy, then $\pi^*(R,$ $T_i)$ satisfies unp$^\#$-privacy.
\end{lem}

\begin{IEEEproof}(Sketch) In Lemma \ref{lem:pipunp}, we have proved that for any $trs \in \{0,1\}^{l_\pi}$, $\pi'(R, T_i)$ satisfies unp$^\#$-privacy.  It means that for any session of $\pi^*$,  the randomness of the transcripts of the $\pi'$'s part are  independent with the transcripts of the $\pi$'s part.
 $\pi'$ is a two-round mutual authentication protocol that each message is authenticated. Then if  both $\pi$ and $\pi'$ achieves unp$^\#$-privacy, $\pi^*(R, T_i)$ satisfies unp$^\#$-privacy.
\end{IEEEproof}

\smallskip
\begin{lem} \label{lem:pistarcu}
If $\pi^*(R, T_i)$ achieves mutual authentication,  and the underlying signature scheme is EU-CMA secure, then $RS^*$ is of PoP credential unforgeability.
\end{lem}

\begin{IEEEproof}(Sketch) We consider two types of adversaries, a dishonest reader $R$ and an external adversary.
%  If the protocol $\pi^*(R, T_i)$ is not of credential unforgeability, then there exists an adversary $\Aa$ who can  $(\epsilon, t, n_1, n_2, n_3, n_4, n_5)$-breaks PoP credential unforgeability of an RFID system $RS^* = (R, \Tt, \func{Setup},$ $\pi^*, \func{CredGen},$ $\func{CredVeri})$.

If the reader $R$ can generate a valid credential $cred = (R, T_i, cred_1, cred_2, cred_3)$ while $cred$ does not correspond to any session between $R$ and $T_i$, then we can construct an algorithm $\Bb$ to pass the EU-CMA experiment using the reader $R$ as a subroutine. Algorithm $\Bb$ sets up an experiment Exp$_{R}^{CredUfrg}$.
$\Bb$ prepares a set of $l$ tags $\Tt$. For each tag $T_j$ ($1 \leq j \leq l$), $\Bb$ selects two random keys $k_{T_j}, k'_{T_j} \in_R \{0,1\}^{l_k}$, and provides them to $R$. Then $\Bb$ selects an index $i$ between $1$ and $l$ randomly.    For each tag $T_j$ ($1 \leq j \leq l$ and $j \neq i$), $\Bb$ selects a random private key $sk_{T_j} \in_R \{0,1\}^{l_{k2}}$ for the signature scheme. $pk$ is the public key provided by the EU-CMA experiment. $\Bb$ sets $T_i$'s public key as $pk$.  $\Bb$ can access an oracle $O_{\Ss_{sk}}$. Given a message $m$, $O_{\Ss_{sk}}$ returns $\Ss_{sk}(m)$, where $sk$ is the matching private key of $pk$. Then $\Bb$ provides all the tags' public keys to the reader $R$. $R$ can query the five oracles. Note that $R$'s queries to $O_1$, $O_3$, and $O_5$ are answered by itself. When $R$ queries $O_2$ regarding to $T_i$, $\Bb$ provides results with $(k_{T_i}, k'_{T_i})$ and the  oracle $O_{\Ss_{sk}}$. If $R$ queries $O_4$ about $T_i$, then $B$ aborts. Finally, when $R$ outputs a PoP credential $cred =(R, T, cred_1, cred_2, cred_3)$ which does not match with any existing  session between $R$ and $T$, if $T = T_i$, $\Bb$ submits $(H(cred_2), cred_3)$ to the EU-CMA experiment. If $R$ can  $(\epsilon, t, n_1, n_2, n_3, n_4, n_5)$-breaks the  PoP credential unforgeability of $RS^*$, then $\Bb$ can break the EU-CMA of the signature scheme with the same advantage $\epsilon$. The running time of $\Bb$ is approximate to that of the reader.

An external adversary $\Aa$ can break the PoP credential unforgeability of $RS^*$ if it can generate a valid credential $cred = \{\Aa, T_i, cred_1, cred_2, cred_3\}$ or $cred = \{R, T_i, cred_1, cred_2, cred_3\}$.  % \{\Aa, T_i, r, \Ss_{sk_\Aa}(r), \Ss_{sk_{T_i}}(H(\Ss_{sk_\Aa}(r)))\}
To generate a credential  $cred = \{\Aa, T_i, cred_1, cred_2, cred_3\}$, adversary   $\Aa$ can choose $r$ randomly, compute $\Ss_{sk_\Aa}(r)$ by itself,  and then set $cred_1= r$ and $cred_2 = \Ss_{sk_\Aa}(r)$.  To get a valid $cred_3$, there are two possible ways. First, adversary $\Aa$ interferes a session of $\pi^*$ between $R$ and $T_i$, blocks the $(2\gamma+1)$-th message $c_{\gamma_1}^* = \{c_{\gamma_1}, c_{\gamma_1}', c_{\gamma_1}''\}$, sends $c_{\gamma_1}^{**} = \{c_{\gamma_1}, H(cred_2), G_{k'_{T_i}}(H(trs||H(cred_2)))\}$ to $T_i$ instead, gets a signature  $\Ss_{sk_{T_i}}(H(cred_2))$ from $T_i$'s reply, and sets it as $cred_3$. If $\Aa$ can generate a valid credential in this way, then $\pi^*(R, T_i)$ does not achieve mutual authentication. Second, adversary $\Aa$  forges a valid  $cred$ without breaking the protocol $\pi^*$. In this case, we can construct an algorithm  $\Bb$  to break the EU-CMA of the underlying signature scheme using $\Aa$ as a subroutine.
Similarly, if adversary $\Aa$ can generate a valid credential  $cred = \{R, T_i, cred_1,cred_2, cred_3\}$, we can construct an algorithm  $\Bb$  to  break the  unp$^\#$-privacy of the protocol $\pi^*$ or the EU-CMA of the underlying signature scheme  using $\Aa$ as a subroutine. \end{IEEEproof}

\section{Proofs for Theorem 4}

\begin{IEEEproof}
According to Theorem 3, $\mathit{MA}$ satisfies adaptive completeness, mutual authentication, and unp$^\#$-privacy. %The Schnorr signature scheme is complete and satisfies EU-CMA \cite{Pointcheval00}, and so as the ECC-based one.
 Thus,  this theorem holds according to Theorem 1   and Theorem  2.\end{IEEEproof}